\documentclass[12pt]{iopart}

\usepackage{iopams}
\usepackage{graphics}
\usepackage{graphicx}
\usepackage[arrow,curve,matrix]{xy}

\newtheorem{theorem}{Theorem}
\newtheorem{lemma}[theorem]{Lemma}
\newtheorem{corrol}[theorem]{Corollary}
\def\QED{\mbox{\rule[0pt]{1.5ex}{1.5ex}}}
\newenvironment{proof}{{\it proof:}}
{\hspace*{\fill}~\QED\par\endtrivlist\unskip}

\def\defn#1{{\bf #1}}

\def\Real{{\mathbb R}}

\def\union{\cup}
\def\Union{\bigcup}
\def\inter{\cap}
\def\Inter{\bigcap}

\def\innerprod(#1,#2){{\left<#1\,,\,#2\right>}}
\def\Set#1{{\left\{#1\right\}}}
\def\qquadtext#1{\qquad\textrm{#1}\qquad}
\def\qquadand{\qquadtext{and}}
\def\quadtext#1{\quad\textrm{#1}\quad}

\def\form#1{{$#1-$}form}
\def\forms#1{{$#1-$}forms}

\def\Man{{\cal M}}

\def\Bman{{\cal B}}
\def\Ebun{{\cal E}}

\def\Id{{\textup{Id}}}
\def\sId{{\textups{Id}}}
\def\dual#1{{\widetilde{#1}}}
\def\Vact(#1){{\left<#1\right>}}
\def\Vacts(#1){{\langle#1\rangle}}
\newsavebox{\ffmul}
\def\Fmul{\,}

\def\Cd{\dot{C}}
\def\Cdd{\ddot{C}}

\def\xdot{{\dot x}}

\def\Cd{{\dot{C}}}

\def\Ebun{{\cal E}}

\def\SigmaHat{{\hat\Sigma}}

\def\Jform{{\cal K}}

\def\probf{{f}}
\def\Jsource{{\cal J}}
\def\pdd{{\Theta}}

\def\textups#1{{\textup{\scriptsize #1}}}

\def\GammaSD{{\Gamma_{\textups{S}}}}
\def\GammaD{{\Gamma_{\textups{D}}}}

\def\supp{{\textup{supp}}}
\def\supp{{\boldsymbol{\cal{S}}}}

\def\deg{{\textup{deg}}}
\def\pr{{\pi}}
\def\varsigmaD{{\boldsymbol\varsigma}}
\def\DD{D}
\def\arhook{\ar@{^(->}}

\def\pfrac#1#2{\frac{\partial #1}{\partial #2}}

\def\LiouV{W}
\def\fna{\textsl{a}}
\def\fnb{\textsl{b}}
\def\fnc{\textsl{c}}
\def\sfna{\textsl{\scriptsize a}}
\def\sfnb{\textsl{\scriptsize b}}
\def\sfnc{\textsl{\scriptsize c}}

\def\ha{h}
\def\metric{g}
\def\zhat{\hat z}
\def\Vy{\underline y}

\begin{document}
\title{Distributional solutions to the Maxwell-Vlasov equations}

\author{Jonathan Gratus}
\address{Physics Department, Lancaster University and
  The Cockcroft Institute.}
\ead{j.gratus@lancaster.ac.uk}

\begin{abstract}
The distributional form of the Maxwell-Vlasov equations are
formulated. Submanifold distributions are analysed and the general
submanifold distributional solutions to the Vlasov equations are
given. The properties required so that these solutions can be a
distributional source to Maxwell's equations are analysed and it is
shown that a sufficient condition is that spacetime be globally
hyperbolic. The cold fluid, multicurrent and water bag models of
charge are shown to be particular cases of the distributional
Maxwell-Vlasov system.
\end{abstract}

\submitto{\JPA}

\section{Introduction}
\label{ch_Intro}

The Maxwell-Vlasov equations give a model for the dynamics of a large
collection of charged particles. They are used to analyse the motion of
beams of particles in a particle accelerator, in order to address
problems such as the effects of coherent synchrotron radiation and
space charge. They are also used for low energy particle dynamics as in
Klystrons and magnetrons. With several species of particles, the
Maxwell-Vlasov equations model the dynamics of plasmas both man made,
as in laser plasma wakefield acceleration and fusion reactors, and
naturally occurring as in the solar winds and the ionosphere.

In this article we formulate the distributional form of the
Maxwell-Vlasov equations, write down the general solution to the
Vlasov equation and identify sufficient conditions such that these 
solutions are valid sources for Maxwell's equations. We then give some
example solutions. This work is relevant in
several areas of active research.

\begin{itemize}
\item
It unifies the Maxwell-Vlasov equations and the cold fluid,
multicurrent and water bag model of charge together with the
Klimontovich distribution.
\item
By comparing the statistical limit of the Klimontovich distribution
with the regular solutions of the Maxwell-Vlasov equations, it may
provide a way of finding a dispersion term for the Boltzmann equation.
\item
In many scenarios, such as micro-bunching and emitance reduction in
accelerators and charge moving on surfaces of constant magnetic flux
in a Tokamak, the resulting charge distribution may be better modelled
by the use of lower dimensional distributions.
\item
It enables the description of the ultra-relativistic expansion for the
Maxwell-Vlasov and the Klimontovich distribution.  Since a single
system applies to all the solutions, finding the expansion of this
system will enable one to write down the expansion of all the
solutions above.
\item
It enables the use of both the retarded Greens potential and the
Li\'{e}nard-Weichart potential to find the electromagnetic field due to an
arbitrary distributional source.
\item
It enables alternative methods of finding numerical results. The
usual method is to use a collection of world-line distributions based
on the motion of individual point particles. Alternative methods could
use ribbon or higher dimensional distributions. This may avoid some of
the regularisation problems associated with point charges.
\end{itemize}

\subsection{Notation}

Sections \ref{ch_Dist} and \ref{ch_Trans} deal with general properties
of distributions and the transport equations (also known as the
Vlasov, Liouville and collisionless Boltzmann equation). For generic
objects in these sections we use the following symbols:

\vspace{1em}
\noindent
\begin{tabular}[t]{@{}p{0.48\textwidth}}
Manifolds: $M,N,P,Q$. 
\\
Subsets: $U,V \subset M$.
\\
Boundary: $\partial M$
\\
Tangent bundle: $TM$.
\\
Bundle of $p-$forms: $\Lambda^p M$.
\\
Bundle of forms: $\Lambda M$.
\\
Generic bundle: $\pi:E\to M$. 
\\
Smooth sections of a bundle: $\Gamma E$.
\\
Space of test forms: $\Gamma_0\Lambda M$.
\\
Space of continuous (not differentiable) \\\quad
forms:
$\Gamma_{\textups{cts}}\Lambda M$.
\\
Space of piecewise continuous \\\quad forms:
$\Gamma_{\textups{pc}}\Lambda M$.
\\
Space of test forms: $\Gamma_0\Lambda M$.
\\
Space of distributions: $\GammaD\Lambda M$.
\\
Set of submanifold distributions: $\GammaSD\Lambda M$.
\\
Vectors and vector fields: $u,v,w$. 
\\
Forms and form fields: $\alpha,\beta$. 
\end{tabular}
\quad
\begin{tabular}[t]{@{}p{0.46\textwidth}}
Evaluation of a field at a point: $\alpha|_x$, $u|_x$
\\
Test forms: $\phi,\psi$.
\\
Distributional forms: $\Psi$, $\Phi$.
\\
Regular distributions: $\DD(\alpha)$.
\\
Degree of a form or distribution:\\\qquad  $\deg(\alpha)$, $\deg(\Psi)$.
\\
Sign involution $\alpha^\eta=(-1)^{\deg(\alpha)}\alpha$.
\\
Smooth map between manifolds: 
\\
\qquad $\fna,\fnb,\fnc$, \quad $\fna\colon N\to M$.
\\
Image set of the map $\fna(N)\subset M$
\\
Preimage set of the subset $U\subset M$: \\\qquad $\fna^{-1}(U)\subset N$
\\
Embedding: $\fna\colon N\hookrightarrow M$.
\\
Composition of maps: $\fna\circ\fnb$.
\\
Pushforward for distributions: $\fna_\varsigma,\fnb_\varsigma$.
\\
Submanifold distribution: $\fna_\varsigmaD\alpha$.
\\
Pullback for distribution: $\fna^\varsigma$.
\end{tabular}
\quad
\begin{tabular}[t]{@{}p{0.46\textwidth}}
Box: $s\colon S\hookrightarrow M$.
\\
Expectation of a distribution $\Psi$ with \\\qquad
respect to box
$s\colon S\hookrightarrow M$: $[s^\varsigma(\Psi)]$.
\end{tabular}
\quad
\begin{tabular}[t]{@{}p{0.46\textwidth}}
Curve: $\gamma\colon\Real\hookrightarrow M$.
\\
Lift of a curve:
$\dot\gamma=\gamma_\star(\partial_\tau)\colon\Real\hookrightarrow T
M$.
\\
Initial hypersurface:
$\sigma\colon\Sigma\hookrightarrow M$.
\end{tabular}

\vspace{1em}

In sections \ref{ch_Intro} and \ref{ch_Examples} we look at the
Maxwell-Vlasov equation on spacetime. We use the following symbols:

\vspace{1em}
\noindent
\begin{tabular}[t]{@{}p{0.35\textwidth}}
Spacetime: $\Man$.
\\
Spacetime metric: $\metric$.
\\
Upper unit hyperboloid: $\Ebun$.
\\
Metric dual of a 1-form: $\dual{\alpha}$.
\\
Electromagnetic 2-form: $F$.
\\
Liouville vector field: $\LiouV$.
\end{tabular}
\quad
\begin{tabular}[t]{@{}p{0.57\textwidth}}
Charge distributional 6-form: $\Theta$.
\\
Electromagnetic source distributional 3-form: $\Jsource$.
\\
Worldline: $C\colon\Real\hookrightarrow\Man$.
\\
Lift of the worldline: $\Cd=C_\star(\partial_\tau)
\colon\Real\hookrightarrow\Ebun$.
\\
Double lift of the worldline: $\Cdd=\Cd_\star(\partial_\tau)
\colon\Real\hookrightarrow T\Ebun$.

\end{tabular}

\subsection{Maxwell-Vlasov equations}
\label{ch_MV}

Let $(\Man,\metric)$ be spacetime and let 
\begin{equation}
\pi\colon \Ebun\to\Man
\qquadtext{where}
\Ebun=\Set{u\in T\Man\,\big|\,\metric(u,u)=-1,\,u^0>0}
\label{MV_Ebun}
\end{equation}
be the unit upper hyperboloid bundle over spacetime. 
Let the coordinates $(x^0,x^1,x^2,x^3,y^1,y^2,y^3)$,
for $\Ebun$ with the embedding $\Ebun\hookrightarrow T\Man$ be
given by $x^a=x^a$, $\xdot^i=y^i$ and let $\xdot^0=y^0=y^0(x^a,y^i)$ be the
solution to $\metric_{ab}y^a y^b=-1$, $y^0>0$, and $y_0=\metric_{0a} y^a$. Here
the indices $a,b,c=0,1,2,3$ and $i,j,k=1,2,3$. We choose physical
units of time and length so that the speed of light $c=1$ and the
permittivity of free space $\epsilon_0=1$.

Given the electromagnetic 2-form field $F\in\Gamma\Lambda^2\Man$, the
standard way of expressing the Maxwell-Vlasov equations is to
prescribe the Vlasov vector field $\LiouV=\LiouV(F)\in\Gamma T\Ebun$, 
which depends on the electromagnetic field as
\begin{equation}
\LiouV=y^a\Fmul \frac{\partial}{\partial x^a}
-
\Gamma^i{}_{bc}\Fmul y^b\Fmul y^c\Fmul \frac{\partial}{\partial y^i}
+
\frac{q}{m} y^a\Fmul F_{ab}\Fmul \metric^{ib}
\frac{\partial}{\partial y^i}
\label{MV_def_W}
\end{equation}
for the species of particle with mass $m$ and charge $q$. For most of
this article we deal with a single species and therefore choose
physical units of mass and charge  so that $m=1$ and $q=1$.  

This vector field is chosen so that is it horizontal, i.e given
$u\in\Ebun$ then
$\pi_\star (\LiouV|_u) = u$
and that if $\gamma\colon \Real\to\Ebun$ is an 
integral curve of $\LiouV$, that is
$\dot\gamma(\tau)=\gamma_\star(\partial_\tau)=\LiouV|_{\gamma(\tau)}$, then
$\gamma=\Cd=C_\star(\partial_\tau)$ 
where $C=\pi\circ\gamma\colon \Real\to\Man$ satisfies the Lorentz
force equation
\begin{equation}
\nabla_{\Cd}\Cd=\dual{i_{\Cd}F}
\label{Lift_force_eqn}
\end{equation}
with $g(\Cd,\Cd)=-1$, see \cite{Yano,Noutchegueme} and lemma
\ref{lm_horizontal}, section \ref{ch_Trans_source}.  Here
$\dual{}:\Lambda^1M\to TM$ is the metric dual given by
$\beta(\dual{\alpha})=g^{-1}(\beta,\alpha)$.  A function
$\probf\in\Gamma\Lambda^0\Ebun$ is called the \defn{one particle
  probability function}.\footnote{The use of the word distribution is
  avoided in this context. This word will be reserved for distribution
  in the sense of Schwartz or De Rham currents.}  The Vlasov equation,
which is also know as the Liouville equation and the collisionless
Boltzmann equation, is now given by
\begin{equation}
\LiouV(\probf)=0
\label{MV_Vlasov_f}
\end{equation}

Maxwell's equations for the electromagnetic field $F$ are given by
\begin{equation}
d F = 0 \qquadand
d \star F = -\Jsource
\label{MV_Maxwell}
\end{equation}
where the source $\Jsource\in\Gamma\Lambda^3\Man$ is
given by
\begin{equation}
\Jsource=
\Big(
\int_{\Real^3}
\frac{\probf y^a}{y_0}\,
\sqrt{|\det(\metric)|}
dy^{123}\Big)
i_{{\partial/\partial x^a}}
\star 1
\label{MV_Maxwell_Curr}
\end{equation}
where $dy^{123}=dy^1\wedge dy^2\wedge dy^3$ etc.
There is a natural non-vanishing 7-form on $\Ebun$ given by 
\begin{equation}
\Omega=-\frac{\det(\metric)}{y_0}
dy^{123}\wedge dx^{0123}
\in\Gamma\Lambda^7\Ebun
\label{MV_def_Omega}
\end{equation}
Since the Lie derivative $L_{\LiouV}\Omega=0$ then
\begin{equation*}
d i_\LiouV (\probf\Omega) = L_\LiouV(\probf\Omega) =
L_\LiouV(\probf)\Omega + \probf L_\LiouV\Omega = 
L_\LiouV(\probf)\Omega = \LiouV(\probf)\Omega 
\end{equation*}
Thus (\ref{MV_Vlasov_f}) is equivalent to $d i_\LiouV(\probf\Omega)=0$.

We introduce the \defn{charge 6-form} given by
$\theta\in\Gamma\Lambda^6\Ebun$ 
\begin{eqnarray}
\theta&=
i_{\LiouV}(\probf\Omega)
\label{MV_theta}
\end{eqnarray}
Thus we can rewrite (\ref{MV_Vlasov_f}) as
\begin{equation}
d\theta = 0
\qquadand
i_\LiouV \theta =0
\label{MV_Vlasov_theta}
\end{equation}
Observe that given $\theta$ and $\Omega$ it is easy to construct the
probability function $\probf$ since the coefficient of $\theta$ with
respect to $dy^{123}\wedge dx^{123}$ is given by $\probf\det(\metric)$.

Comparing (\ref{MV_Maxwell_Curr}) and (\ref{MV_theta}) we see that the 
integration for the source
$\Jsource\in\Gamma\Lambda^3\Man$, is in fact a integration along a
fibre \cite{BottTu} given by
\begin{equation}
\int_\Ebun \pi^\star\phi\wedge\theta
=
\int_\Man \phi\wedge\Jsource
\label{MV_int_theta}
\end{equation}
for all test 1-forms $\phi\in\Gamma_0\Lambda^1\Man$, that is
the space of smooth 1-forms with compact support on $\Man$.

The 6-form $\theta$ has a natural interpretation in terms of
probabilities or charge.
Given a ``box'', that is a compact 6-dimensional region
$S\subset\Ebun$ which is transverse to $\LiouV$ then the probability that
the particle passes though $S$ (or the total charge passing though $S$)
is given by the integral $\int_S\theta$.
An example of such a box is the set 
$S=\Set{(x^a,y^i)|x^0=x^0_c\,,\ x^i_l<x^i<x^i_u\,,\ y^i_l<y^i<y^i_u}$
for some constants $\Set{x^0_c,x^i_l,x^i_u,y^i_l,y^i_u}$. Assuming
the correct orientation for $S$ it is usual to demand
$\int_S\theta\ge0$. An initial hypersurface $\Sigma\subset\Ebun$ is any
6 dimensional hypersurface transverse to $\LiouV$. If $\theta$ is to be
interpreted as  probability then we further demand that
$\int_\Sigma\theta=1$. However if $\theta$ is the charge distribution
then $\int_\Sigma\theta=Q_{\textups{total}}$ is the total charge. It is
easy to show that $Q_{\textups{total}}$ is a conserved quantity.
In order to evaluate a distribution on a box or a hypersurface we
require the definition of the pullback of a distribution given in
section \ref{ch_Pull}.

\vspace{1em}

In order to write the Maxwell-Vlasov system in distributional language
we replace the 6-form $\theta$ with the \defn{charge distributional
  6-form}
\begin{equation}
\pdd\in\GammaD\Lambda^6\Ebun
\,,
\label{MV_Liou_field}
\end{equation}
This is defined as an element in the
dual to the space of test 1-forms $\Gamma_0\Lambda^1\Ebun$. Given a
test 1-form $\phi\in\Gamma_0\Lambda^1\Ebun$ then the action of
$\Theta$ on $\phi$ is the real number written $\Theta[\phi]$. 

Inspired by (\ref{MV_int_theta}) we define the pushforward of a distribution
$\pi_\varsigma$ \cite{DeRham} via
\begin{equation}
\pi_\varsigma(\Theta)[\phi]=\Theta[\pi^\star(\phi)]
\label{MV_pi_Theta}
\end{equation}
and we have the source for Maxwell's equations as
\begin{equation}
\Jsource \in\GammaD\Lambda^3\Man
\;;\qquad
\Jsource = \pi_\varsigma \pdd
\label{MV_def_Jsource}
\end{equation}
The details of this are given in section \ref{ch_Push}

Since $\Jsource$ is a distribution, then in general, the solution to
Maxwell's equations (\ref{MV_Maxwell}) will also be a
distribution. However in order for $F$ to drive the Liouville
equation (\ref{MV_def_W},\ref{MV_Vlasov_theta}) we require that $F$ is
continuous, though not necessarily differentiable.
Since $F$ is not differentiable, we must convert it into a
distribution before inserting it into Maxwell's equations. Every
continuous $q$-form $\alpha\in\Gamma_{\textups{cts}}\Lambda^q M$ 
gives rise to a regular distribution $q$-form 
$D(\alpha)\in\GammaD\Lambda^q M$.
Thus we can define $D(F)\in\GammaD\Lambda^2\Man$ given by
$\DD(F)[\phi] = \int_\Man \phi\wedge F$ for any
$\phi\in\Gamma_0\Lambda^{2}\Man$. Maxwell's equations become
\begin{equation}
d (\DD{F}) = 0
\qquadand
d(\DD (\star F)) = -\Jsource
\label{MV_Maxwell_dist}
\end{equation}
As is noted in section \ref{ch_Dist} we can define the exterior
derivative and the internal contraction of a distribution, thus the
Liouville equation (\ref{MV_Vlasov_theta}) becomes the equivalent
for $\Theta$ namely
\begin{equation}
d \pdd = 0 
\label{MV_pdd_closed}
\end{equation}
and
\begin{equation}
i_\LiouV \pdd =0
\label{MV_Liou_eqn}
\end{equation}
where the \defn{Liouville operator} $\LiouV$ is given by
(\ref{MV_def_W}). These are also called the \defn{transport equations}
of $\Theta$ for $\LiouV$.  Theorem \ref{thm_Transp_source_closed}
implies that the source $\Jsource$ is closed, $d\Jsource=0$.

We limit ourselves to ``submanifold distributions''. These are written
$\Theta=\fna_\varsigmaD\alpha$ where $\fna\colon N\hookrightarrow
\Ebun$ is an an embedding with $N$ possibly having a boundary and
$\alpha\in\Gamma\Lambda^p N$ is a smooth $p-$form where $p=\dim N-1$.
The action of this distribution on a test form
$\phi\in\Gamma_0\Lambda^1\Ebun$ is given by
\begin{equation}
\fna_\varsigmaD\alpha[\phi]=\int_N \fna^\star(\phi)\wedge \alpha
\label{MV_def_Push}
\end{equation}
Most of section \ref{ch_Dist}  details the
mathematics needed to manipulate these distributions.

The geometric nature of these distributions means that we can define
many useful properties such as pullbacks, without having the worry
about the convergence and other topological properties. However these
distributions are sufficiently general that physically interesting
solutions to the Maxwell-Vlasov system can be written using them.  In
section \ref{ch_Examples} we show that the cold fluid, the
multicurrent and the water bag models of charge dynamics are all
examples of submanifold distributional solutions to the Maxwell-Vlasov
equations. 
Another example of a submanifold distribution, discussed in section
\ref{ch_point}, is the worldline for a single
charged particle given by $\pdd=\Cd_\varsigmaD(1)$ so that
$\pdd[\phi]= \int_{\Real} \Cd^\star (\phi)$ for all
$\phi\in\Gamma_0\Lambda^1\Ebun$. In Minkowski space, this is equivalent to the
usual expression for a charged particle, written in terms of a Dirac
$\delta$-function probability function as
\begin{equation*}
f=\Big(\prod_{i=1}^3\delta(x^i-C^i(\tau))\Big)\Big(
\prod_{j=1}^3\delta(y^j-\Cd^j(\tau))\Big)
\end{equation*}
where $t=C^0(\tau)$ so that $\tau=(C^0)^{-1}(t)$. Thus we have
\begin{equation}
\Cd_\varsigmaD(1)=i_\LiouV(f\Omega)
\label{MV_point_equiv}
\end{equation}
To see this recall that $\Cd:\Real\to\Ebun$ is an integral curve of
$\LiouV$ so that $\LiouV|_\Cd(\tau)=\Cdd(\tau)$ where
$\Cdd=\Cd_\star(\partial_\tau)$, thus given a test 1-form
$\phi\in\Gamma_0\Lambda^1\Ebun$ we have
\begin{equation*}
\fl
\Cd^\star\phi|_{\tau}
=\!
i_{\partial_\tau}(\Cd^\star\phi)|_{\tau} d\tau
=\!
\Cd^\star( i_{\Cd_\star(\partial_\tau|_\tau\!)}\phi ) d\tau
=\!
\Cd^\star( i_{\Cdd(\tau)}\phi ) d\tau
=\!
\Cd^\star( i_{\LiouV|_{\Cd(\tau)}}\phi ) d\tau
=\!
( i_{\LiouV}\phi )|_{\Cd(\tau)} d\tau
\end{equation*}
Hence
\begin{equation*}
\Cd_\varsigmaD(1)[\phi]=
\int_\Real \Cd^\star(\phi)=
\int_\Real ( i_{\LiouV}\phi )|_{\Cd(\tau)} d\tau
\end{equation*}
and
\begin{eqnarray*}
\fl\qquad
i_\LiouV(f\Omega)[\phi]
&=
(f\Omega)[i_\LiouV\phi]
\\\fl&=
\int 
\Big(\prod_{i=1}^3\delta(x^i-C^i(\tau))\Big)
\Big(\prod_{j=1}^3\delta(y^j-\Cd^j(\tau))\Big)
(i_\LiouV\phi)
\frac{dx^{123}\wedge dy^{123} \wedge dt}{y^0}
\\\fl&=
\int_\Real
(i_\LiouV\phi)|_{\Cd(\tau)}
\frac{1}{\Cd^0(\tau)}
dt 
=
\int_\Real ( i_{\LiouV}\phi )|_{\Cd(\tau)} \frac{d\tau}{d t} dt
=
\int_\Real ( i_{\LiouV}\phi )|_{\Cd(\tau)} d\tau
\end{eqnarray*}
Equation (\ref{MV_point_equiv}) is true for general spacetimes,
assuming the correct definition of the \mbox{$\delta$-function}. One of the
advantages of the pushforward approach to distributions is that the
definitions do not explicitly contain the coordinates system. They also
do not depend on the metric and therefore can be applied to the upper
unit hyperboloid bundle $\Ebun$ without having to choose a metric for
this manifold.

The Klimontovich distribution is given by the sum of worldline
distributions one for each particle \cite{davidson}.  We show that the
worldline of a point charge and the Klimontovich distribution satisfy
the Liouville equation and have the correct form for the source of
Maxwell's equations. However the solutions to Maxwell's equations with
the worldline as a source, which in free space are given by the
Li\'{e}nard-Weichart potentials, are not continuous and therefore we
cannot insert them into the Liouville equation without some form of
regularisation, leading, for example, to the Lorentz-Dirac
equation. The same is true for the Klimontovich distribution, where
the usual regularisation is simply to ignore the field due to a
particular particle when looking at the dynamics of that particle.

\begin{figure}
\centerline{
\includegraphics[width=0.5\textwidth,viewport=66 143 477 519]{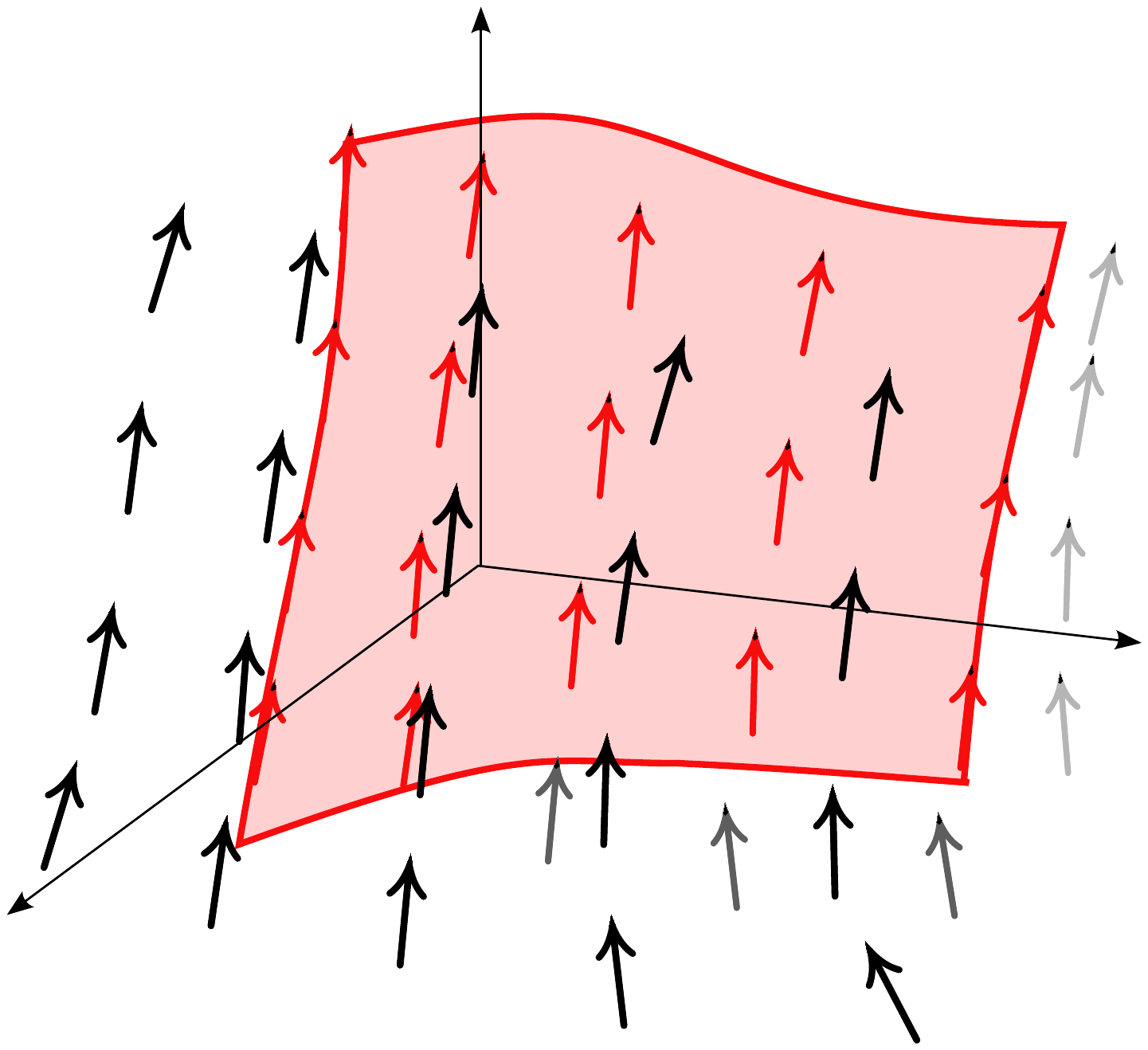}}
\caption{A submanifold solution to the transport equation. Observe
  that the submanifold is tangential to the vector field and that the
  boundary of submanifold is also tangential to the vector field.}
\label{fig_soln}
\end{figure}

In section \ref{ch_Trans} we look at the submanifold distributional
solutions to the transport equations for a general vector field
$v\in\Gamma TM$ on a general manifold $M$. Theorem \ref{lm_Trans_soln}
gives the necessary and sufficient conditions for a submanifold
solution. A sketch of such a solution is given in figure
\ref{fig_soln}.
We look at initial hypersurfaces and boxes and show the
conservation of charge, theorem \ref{lm_Trans_S12}, as well as that,
given a distribution on an initial hypersurface, there
exists a unique solution to the transport equations, theorem
\ref{thm_Transp_initial_distr}. This is achieved by extending the
initial value along the integral curves of $v$.

Given a solution $\Theta$ to the Liouville equations on $\Ebun$ we can
ask if $\Jsource=\pi_\varsigma\Theta$ is a valid source for Maxwell's
equations. The problem is that, in general, $\pi_\varsigma\Theta$ may
not be a distribution. This is because $\pi$ is not a proper map. A
map is \defn{proper} if the preimage of any compact set is
compact. This is required so that $\pi^\star(\phi)$ in
(\ref{MV_pi_Theta}) has compact support.  In section
\ref{ch_Trans_source}, lemma \ref{lm_Trans_source_bdd}, we show that
an appropriately bounded submanifold distribution is a valid
source. We say that \defn{$\Theta$ is bounded with respect to $\pi$}.
This is the case in the examples given in section \ref{ch_Examples}.
We then pose the following question (\ref{Trans_source_question}): If
a distribution is initially bounded, does it remain bounded?
Surprisingly the answer to this question is no and two counter
examples are given, one using a drifting source in Galilean spacetime,
the other, on Minkowski spacetime, where the initial data is specified
on a lightlike surface. However if, as we generally assume, $\Man$ is
a globally hyperbolic spacetime, and the initial data is given on a
Cauchy surface, then theorem \ref{thm_Transp_source_init} shows that
the solution will remain bounded. However, as in the case of the
worldline for a point charge, the source may be valid distribution but
the resulting electromagnetic 2-form need not be continuous.

\vspace{1em}
It is easy to extend (\ref{MV_def_Jsource}-\ref{MV_Liou_eqn}) to model
several species of particle with masses $m_\alpha$ and charges
$q_\alpha$. Let $\pdd_\alpha\in\GammaD\Lambda^6\Ebun$ be the
corresponding one particle 6-form distribution, and $\LiouV_\alpha\in\Gamma
T\Man$ be the Vlasov vector field (\ref{MV_def_W}) with appropriate
$m$ and $q$. The Vlasov equations are given by
\begin{equation}
d\pdd_\alpha = 0\qquadand i_{\LiouV_\alpha}\pdd_\alpha = 0
\label{MV_Liou_eqn_alpha} 
\end{equation}
and Maxwell's equations are given by (\ref{MV_Maxwell_dist}) where
the source 
\begin{equation}
\Jsource=\sum_\alpha \pi_\varsigma(\pdd_\alpha)
\label{MV_Maxwell_dist_alpha}
\end{equation}


\section{Distributions}
\label{ch_Dist}

Let $M$ be an arbitrary manifold of dimension $m$. 
Let $\Gamma_0\Lambda^p M$ be the
space of \defn{test $p$-forms.} 
\begin{equation}
\Gamma_0\Lambda^p M=\Set{\phi\in\Gamma\Lambda^p M\,\Big|\,\phi\ \textup{has
    compact support}}
\label{Dist_Test_Man}
\end{equation}
Let the space of \defn{$p$-forms distributions} be the vector space
dual of the space of test functions.
\begin{equation}
\GammaD\Lambda^p M=
\big(\Gamma_0\Lambda^{m-p}M\big)^\star
\label{Dist_DistMan}
\end{equation}
We use square bracket notation.
\begin{equation}
\GammaD\Lambda^{m-p} M\times\Gamma_0\Lambda^p M\to\Real\,,\qquad
(\Psi,\phi)\mapsto \Psi[\phi]\in\Real
\label{Dist_distr_notat}
\end{equation}
Observe that the space of piecewise continuous \forms{p} is a subspace
of the space of \form{p} distributions. These are called \defn{regular
  distributions}.
\begin{equation}
\fl\quad
\DD\colon \Gamma_{\textups{pc}}\Lambda^{p}M\hookrightarrow \GammaD\Lambda^{p}M
\;;\qquad
\DD(\alpha)[\phi] = 
\int_M \phi\wedge\alpha
\qquadtext{for}\phi\in\Gamma_0\Lambda^{m-p}M
\label{Dist_def_D_alpha}
\end{equation}
The \defn{exterior derivative} of a \form{p} distribution is given by
\begin{equation}
\fl\quad
d\colon \GammaD\Lambda^p M\to\GammaD\Lambda^{p+1}M
\;;\qquad
d\Psi[\phi]=-\Psi[d\phi^\eta]
\qquadtext{for}
\phi\in\Gamma_0\Lambda^{m-p+1}M
\label{Dist_exterior_d}
\end{equation}
The \defn{internal contraction} of a vector field 
$v\in\Gamma TM$ with a $p-$form
distribution $\Psi\in\GammaD\Lambda^p M$ is given by
\begin{eqnarray}
\Gamma TM\times\GammaD\Lambda^p M\to\GammaD\Lambda^{p-1}M
\,,\quad
(v,\Psi)\mapsto i_v\Psi
\label{Dist_internal_contr}
\\
\quadtext{where}
i_v\Psi[\phi] = 
-\Psi[i_v\phi^\eta]
\quadtext{for all}
\phi\in\Gamma_0\Lambda^{m-p+1}M
\nonumber
\end{eqnarray}
Given a regular distribution $\DD(\beta)$ with 
$\beta\in\Gamma_{\textups{pc}}\Lambda M$
then
\begin{equation}
i_v (\DD\beta) = \DD(i_v \beta)
\end{equation}
Given a regular distribution $\DD(\beta)$ with
$\beta\in\Gamma_{\textups{cts}}\Lambda M$ and $\beta$ has a continuous
derivative and where $M$ has no boundary then
\begin{equation}
d (\DD\beta) = \DD(d\beta)
\label{Dist_D_basic_res}
\end{equation}

We define the \defn{support of a distribution} $\supp(\Psi)\subset M$ as
\begin{equation}
\fl\quad
\supp(\Psi)=
\Inter\Set{U\!\subset\! M\,\Big|\,U\textup{ is closed and if }
\phi\in\Gamma_0\Lambda M,\
\supp(\phi)\!\subset\! M\backslash U\textup{ then }
\Psi[\phi]=0}
\label{Dist_supp}
\end{equation}
where $\supp(\phi)$ is the closed support of $\phi$.
That is $\supp(\Psi)$ is the smallest closed subset of $M$ such that
for every test form $\phi$ with
$\supp(\Psi)\inter\supp(\phi)=\emptyset$ then $\Psi[\phi]=0$.

\begin{lemma}
\label{lm_Dist_reg_supp}
For regular distributions
$\supp(\DD(\alpha))=\supp(\alpha)$.
\end{lemma}
\begin{proof}
Clearly if $\supp(\phi)\inter\supp(\alpha)=\emptyset$ then 
$\DD(\alpha)[\phi]=\int\phi\wedge\alpha=0$. Thus
$\supp(\DD(\alpha))\subset\supp(\alpha)$. If 
$\supp(\DD(\alpha))\ne\supp(\alpha)$ then there exists an open set
$U\subset\supp(\alpha)\backslash\supp(\DD(\alpha))$. It is easy to
construct a test form $\phi$ with $\supp(\phi)\subset U$ such that
$\int\phi\wedge\alpha\ne0$. So
$\supp(\DD\alpha)\inter U\ne 0$
leading to the contradiction $\supp(\DD\alpha)\inter
\big(\supp(\alpha)\backslash\supp(\DD\alpha))\ne\emptyset$.
Hence result. 
\end{proof}

\begin{lemma}
For $\Psi\in\GammaD\Lambda M$ and $v\in\Gamma T M$
\begin{equation}
\supp(d\Psi)\subset \supp(\Psi)
\qquadand
\supp(i_v\Psi)\subset \supp(\Psi)
\label{Dist_supp_d_iv}
\end{equation}
\end{lemma}
\begin{proof}
Given $\phi\in\Gamma_0\Lambda M$ such that $\supp(\phi)\subset
M\backslash\supp(\Psi)$ then $\supp(d\phi)\subset
M\backslash\supp(\Psi)$ and $\supp(i_v\phi)\subset
M\backslash\supp(\Psi)$. Thus 
$0=-\Psi[d\phi^\eta]=d\Psi[\phi]$ and
$0=-\Psi[i_v\phi^\eta]=i_v\Psi[\phi]$. Hence
$M\backslash\supp(\Phi)\subset M\backslash\supp(d\Phi)$ and 
$M\backslash\supp(\Phi)\subset M\backslash\supp(i_v\Phi)$ and
(\ref{Dist_supp_d_iv}) follows.
\end{proof}


\subsection{Pushforward of distributions}
\label{ch_Push}

The \defn{pushforward} \cite{DeRham} of a distribution with respect to
a smooth map $\fna\colon N\to M$ where $\dim M=m$ and $\dim N=n$ is
given by
\begin{equation}
\fl\qquad
\fna_\varsigma \colon \GammaD \Lambda^{p} N \to \GammaD \Lambda^{m-n+p} M
\;;\qquad
\fna_\varsigma(\Psi)[\phi] = \Psi[\fna^\star(\phi)] 
\quadtext{for}
\phi\in\Gamma_0 \Lambda^{n-p} M
\label{Push_def}
\end{equation}
The pushforward does not preserve the degree of a distribution. Instead
we have 
\begin{equation}
\deg(\fna_\varsigma(\Psi))=\deg(\Psi)+\dim M - \dim N
\label{Push_deg}
\end{equation}

We note that the pushforward given in definition (\ref{Push_def}) is
not always defined. The problem is that in general there is no guarantee
that $\fna^\star(\phi)$ has compact support. This can lead to the 
pushforward being infinite as in the following example. 
Let $\fna\colon \Real\to\Set{x}$ and let
$\Psi=\DD(d t)\in\GammaD\Lambda^{1}\Set{x}$. For
$\phi\in\Gamma_0\Lambda^0\Set{x}$ we have
\begin{equation*}
\fna_\varsigma(\Psi)[\phi]=\Psi[\fna^\star(\phi)]=\int_\Real \phi(x)\Fmul d
t=\phi(x)\Fmul \int_\Real d t
\end{equation*}
which is undefined if $\phi(x)\ne0$. It can also
lead to problems with boundaries.

By contrast, we can define the pushforward of a distribution when the map
$\fna\colon N\to M$ is \defn{proper}, that is, when the preimage of every
compact set is compact. However the bundle map $\pi\colon\Ebun\to\Man$
is not proper but from (\ref{MV_pi_Theta}) we see we wish to define
the pushforward of a distribution with respect to $\pi$. This is not
always possible, but it is for certain distribution which we say are
bounded with respect to $\pi$. 

We say the distribution
$\Psi\in\GammaD\Lambda N$ \defn{is bounded with respect to}
$\fna\colon N\to M$ if
for any compact set $U\subset M$ then
$\supp(\Psi)\inter\fna^{-1}(U)\subset N$ is compact. 
In this case we
define
\begin{equation}
\fna_\varsigma(\Psi)[\phi] = \Psi[h\,\fna^\star(\phi)]
\qquadtext{for}
\phi\in\Gamma_0\Lambda M
\label{Push_def_bdd}
\end{equation}
where $h\in\Gamma_0\Lambda^0 N$ is any test bump function such that
\begin{equation}
\supp(\Psi)\inter\supp(1-h)\inter\supp(\fna^\star\phi)=\emptyset
\label{Push_bdd_h_req}
\end{equation}
\begin{lemma}
The pushforward of a distribution bounded with respect to $\fna$ is
well defined.
\end{lemma}
\begin{proof}
Since $\supp(\phi)$ is compact and
$\supp(\fna^\star\phi)\subset\fna^{-1}(\supp(\phi))$ then 
$\supp(\fna^\star\phi)\inter\supp(\Psi)$ is compact. There exists a
compact set $V_1$ and $V_2$ such that
$\supp(\fna^\star\phi)\inter\supp(\Psi)\subset\textup{interior}(V_1)$
and $V_2\subset\textup{interior}(V_1)$ and a bump function $h$ such that
$h(x)=1$ for $x\in V_1$ and $h(x)=0$ for $x\notin V_2$. 

To see that $\fna_\varsigma(\Psi)[\phi]$ is well defined, consider two
functions $h$ and $\hat{h}$, then since $h-\hat{h}=(1-\hat{h}) -
(1-h)$ we have $\supp(h-\hat{h})\subset
\supp(1-h)\union\supp(1-\hat{h})$. Thus
\begin{eqnarray*}
\fl\qquad\lefteqn{
\supp(\Psi)\inter \supp(\fna^\star\phi) \inter\supp(h-\hat{h})
}\qquad&
\\
&\subset
\Big(\supp(\Psi)\inter \supp(\fna^\star\phi) \inter\supp(1-\hat{h})\Big)
\union
\Big(\supp(\Psi)\inter \supp(\fna^\star\phi) \inter\supp(1-\hat{h})\Big)
=
\emptyset
\end{eqnarray*}
Thus $\Psi[(h-\hat{h})\,\fna^\star(\phi)]=0$.
\end{proof}

To guarantee that the pushforward is well defined we limit ourselves to
three cases: 
\begin{itemize}
\item
When $\fna\colon N\to M$ is proper.  
\item
When $\Psi$ is bounded with respect to $\fna$.
\item
When $\pi\colon E\to M$ is a fibre bundle and
$\alpha\in\Gamma\Lambda{E}$ has the property that integration along
the fibres $\pi^{-1}(x)$, given by $\int_E \pi^\star\phi\wedge
(\pi_\varsigma(\DD\alpha)) = \int_M \phi\wedge\alpha$ for all $\phi$
is defined \cite{BottTu}. This last case is simply for regular
distributions which are not bounded with respect to $\pi$ but such
that integral exists. We will not consider this standard type of
pushforward further in the article.
\end{itemize}

\begin{lemma}
\label{lm_Push_composition}
Given manifolds $N,M,P$ and smooth proper maps $\fnb\colon P\to N$ and 
$\fna\colon N\to M$ then the composition of the push forwards is the push forward of
the composition:
\begin{equation}
(\fna\circ \fnb)_\varsigma=\fna_\varsigma\circ \fnb_\varsigma
\label{Push_composition}
\end{equation}
\end{lemma}
\begin{proof}
The composition of two proper maps is proper.
Let $\Psi\in\GammaD\Lambda M$ and $\phi\in\Gamma_0\Lambda P$ then
\begin{equation*}
\fna_\varsigma(\fnb_\varsigma(\Psi))[\phi]
=
\fnb_\varsigma(\Psi)[\fna^\star\phi]
=
\Psi[\fnb^\star \fna^\star\phi]
=
\Psi[(\fna\circ \fnb)^\star \phi]
=
((\fna\circ \fnb)_\varsigma\Psi)[\phi]
\end{equation*}
\end{proof}
Observe that if $\fna$ or $\fnb$ is not proper then care must be
taken. See lemma \ref{lm_Trans_source_bdd}.

\begin{lemma}
\label{lm_Push_d_comm}
If $\fna\colon N\to M$ is proper then the pushforward of distributions
commutes with the exterior derivative.
\begin{equation}
d\circ \fna_\varsigma = \fna_\varsigma \circ d
\label{Push_comm_d}
\end{equation}
\end{lemma}
\begin{proof}
\begin{eqnarray*}
\fl\quad
d (\fna_\varsigma \Psi) [\phi]
&=
- \fna_\varsigma \Psi [d\phi^\eta]
=
- \Psi [\fna^\star(d\phi^\eta)]
=
- \Psi [d \fna^\star(\phi^\eta)]
=
- \Psi [d (\fna^\star\phi^\eta)]
=
d \Psi [\fna^\star(\phi)]
\\\fl&=
\fna_\varsigma (d \Psi) [\phi]
\end{eqnarray*}
\end{proof}

\begin{lemma}
\label{lm_Push_d_comm_bdd}
Let $\fna\colon N\to M$ be smooth  and let $\Psi$ be bounded with
respect to $\fna$, then 
\begin{equation}
d (\fna_\varsigma \Psi)=\fna_\varsigma (d\Psi)
\label{Push_comm_d_bdd}
\end{equation}
\end{lemma}
\begin{proof}
From (\ref{Dist_supp_d_iv}), $d\Psi$ is bounded with respect to
$\fna$. Given $h\in\Gamma_0\Lambda^0 N$ such that
(\ref{Push_bdd_h_req}) holds, then from (\ref{Dist_supp_d_iv}) 
$\supp(d\Psi)\inter\supp(1-h)\inter\supp(\fna^\star\phi)=\emptyset$
and since $\supp(dh)\subset\supp(1-h)$ we have
$\supp(\Psi)\inter\supp(dh)\inter\supp(\fna^\star\phi)=\emptyset$ so
$\Psi[dh\wedge\fna^\star\phi^\eta]=0$. Thus
\begin{eqnarray*}
\fl\quad
d (\fna_\varsigma \Psi) [\phi]
&=
- \fna_\varsigma \Psi [d\phi^\eta]
=
- \Psi [h\,\fna^\star(d\phi^\eta)]
=
- \Psi [h\,d \fna^\star(\phi^\eta)]
\\\fl&
=
- \Psi [d (h\,\fna^\star\phi^\eta)]
+ \Psi[dh\wedge\fna^\star\phi^\eta]
=
- \Psi [d (\fna^\star\phi^\eta)]
=
(d \Psi) [\fna^\star(\phi)]
=
\fna_\varsigma (d \Psi) [\phi]
\end{eqnarray*}
\end{proof}

Given $\fna\colon N\to M$ and a vector field we say that $v\in\Gamma T M$
\defn{is tangential to} $\fna$ if there
exists $u\in\Gamma T N$ such that for each $x\in N$,
$\fna_\star(u|_x)=v|_{\fna(x)}$. We write $\fna_\star(u)=v|_{\fna(N)}$ if 
$\fna_\star(u|_x)=v|_{\fna(x)}$ for all $x\in N$,

\begin{lemma}
\label{lm_Push_iv_comm}
Let $\fna\colon N\to M$ be proper with $v\in\Gamma T M$ tangential to
$\fna$ and let $u\in\Gamma T N$ satisfy
$\fna_\star(u)=v|_{\fna(N)}$ then
\begin{equation}
i_v \circ \fna_\varsigma
=
\fna_\varsigma \circ i_{u}
\label{Push_v_tanj}
\end{equation}
\end{lemma}
\begin{proof}
Given $\Psi\in\GammaD\Lambda M$, 
$\phi\in\Gamma_0\Lambda M$, $\fna\colon N\to M$ and $x\in N$, let
$y=\fna(x)$ then
\begin{equation*}
\fl\quad
(i_u \fna^\star\phi)|_{x}
=
i_{u|_x} \fna^\star_x (\phi|_{y})
=
\fna^\star_x \big(i_{\fna_\star(u|_x)} (\phi|_{y}) \big)
=
\fna^\star_x \big( i_{v|_{y}} (\phi|_{y})\big)
=
\fna^\star_x \big((i_v\phi)|_{y}\big)
=
\big(\fna^\star (i_v\phi)\big)|_{x}
\end{equation*}
where $\fna^\star_x\colon \Lambda_y N\to\Lambda_x M$.
Thus $i_u \fna^\star\phi=\fna^\star (i_v\phi)$.
Hence
\begin{equation*}
\fl\quad
i_v \fna_\varsigma(\Psi)[\phi]
=
-\fna_\varsigma(\Psi)[i_v\phi^\eta]
=
-\Psi[\fna^\star (i_v\phi^\eta)]
=
-\Psi[i_u \fna^\star\phi^\eta]
=
i_u \Psi[\fna^\star\phi]
=
\fna_\varsigma (i_u \Psi)[\phi]
\end{equation*}
\end{proof}

\begin{lemma}
\label{lm_Push_supp}
Let $\fna\colon N\to M$ be proper and $\Psi\in\GammaD\Lambda N$ then
\begin{equation}
\supp(\fna_\varsigma\Psi)\subset \fna(\supp\,\Psi)
\label{Push_supp}
\end{equation}
\end{lemma}
\begin{proof}
Given any $x\in \fna^{-1}(M\backslash \fna(\supp\,\Psi))$ then 
$\fna(x)\in M\backslash \fna(\supp\,\Psi)$ and so $\fna(x)\notin
\fna(\supp\,\Psi)$ so $x\notin \supp(\Psi)$
hence $x\in N\backslash\supp(\Psi)$. This gives
$\fna^{-1}(M\backslash \fna(\supp\,\Psi))\subset N\backslash\supp(\Psi)$

Let $\phi\in\Gamma_0\Lambda M$ with $\supp(\phi)\subset M\backslash
\fna(\supp\,\Psi)$. Then
\begin{equation*}
\supp(\fna^\star\phi)=
\fna^{-1}(\supp\,\phi)
\subset \fna^{-1}(M\backslash \fna(\supp\,\Psi))
\subset N\backslash\supp(\Psi)
\end{equation*}
Thus $\fna_\varsigma\Psi[\phi]=\Psi[\fna^\star\phi]=0$. 

Hence we have shown $\supp(\phi)\subset M\backslash \fna(\supp\,\Psi)$
implies $\fna_\varsigma\Psi[\phi]=0$. Hence $M\backslash
\fna(\supp\,\Psi)\subset M\backslash\supp(\fna_\varsigma\Psi)$ and hence
(\ref{Push_supp}).

\end{proof}

To see that in general $\supp(\fna_\varsigma\Psi)\ne \fna(\supp\,\Psi)$
consider the following counter example. Let $M=\Real$ and
$N=\Real\times\Set{-1,1}$ with $\fna\colon N\to M$ given by $\fna(x,i)=x$. Let
$\Psi=D(\alpha)\in\GammaD\Lambda N$ where
$\alpha\in\Gamma\Lambda^0 N$ is given by
$\alpha|_{(x,1)}=1$ and 
$\alpha|_{(x,-1)}=-1$
then given $\phi\in\Gamma_0\Lambda^{1} M$ we have
\begin{equation*}
\fl\quad
\fna_\varsigmaD\alpha[\phi]=
\int_{N}\fna^\star(\phi)\wedge\alpha=
\int_{x\in\Real} \phi|_{x}\wedge\alpha|_{(x,1)}+
\int_{x\in\Real} \phi|_{x}\wedge\alpha|_{(x,-1)}
=
\int_{\Real} \phi-
\int_{\Real} \phi
=
0
\end{equation*}
Hence $\supp(\fna_\varsigma\Psi)=\emptyset\ne \Real=\fna(\supp\,\Psi)$.

\subsection{Submanifold Distributions}
\label{ch_SMD}

A special kind of distribution is the submanifold distribution. Before
we introduce submanifold distributions we observe some basic
properties of closed embedding, i.e. embedding $\fna\colon
N\hookrightarrow M$ where $\fna(N)\subset M$ is closed. First observe
that $\fna$ is proper.
\begin{lemma} 
\label{lm_SMD_fna_proper}
Given an embedding $\fna\colon N\hookrightarrow M$ such that 
$\fna(N)\subset M$ is closed it follows that $\fna$ is proper.
\end{lemma}
\begin{proof}
Given a compact $U\subset M$ then $\fna(N)\inter U$ is closed and
therefore compact. The restriction $\fna|_{\fna^{-1}(U)}\colon
\fna^{-1}(U)\to \fna(N)\inter U$ is a homeomorphism so $\fna^{-1}(U)$
is compact. Thus $\fna$ is proper.
\end{proof}

The next three lemmas use the embedding nature of $\fna$ to relate
$\fna_\varsigma(\DD\alpha)$ and $\alpha$.
\begin{lemma}
\label{lm_embed_fna_star_inv}
Let $\fna\colon N\hookrightarrow M$ be an embedding. Given an open set $U\subset M$ and
$\phi\in\Gamma\Lambda N$ such that $\supp(\phi)\subset \fna^{-1}(U)$ then there
exists $\psi\in\Gamma\Lambda M$ such that $\fna^\star\psi=\phi$ and
$\supp(\psi)\subset U$.
\end{lemma}
\begin{proof}
Consider $x\in \fna(N)$. Since $\fna$ is an embedding there exists an
open coordinate patch $V\subset M$, $(x^1,\ldots,x^m)$ about $x$ and a
coordinate patch $\fna^{-1}V\subset N$, $(y^1,\ldots,y^n)$ such that
$\fna(y^1,\ldots,y^n)=(x^1,\ldots,x^n,0,\ldots,0)$.

Assume for the moment that $\supp(\phi)\subset \fna^{-1}(V)$ and
$\phi=\sum_I \phi_I dy^I$ where $I\subset\Set{1,\ldots n}$ and
$dy^I=dy^{I_1}\wedge\cdots\wedge dy^{I_{|I|}}$
refers to multi-index notation, then let
\begin{equation*}
\psi=\sum_{I\subset\Set{1,\ldots n}} \phi_I dx^I \ha(x^{n+1},\ldots,x^m)
\end{equation*}
where $\ha(x^{n+1},\ldots,x^m)$ is a smooth function with
$\ha(0,\ldots,0)=1$ and so that $\supp(\psi)\subset V$.
Thus $\fna^\star\psi=\phi$.

Now in general use a partition of unity to partition $U$ into
coordinate patches $V_i$. Thus if $\phi=\sum \phi_i$ with
$\supp(\phi_i)\subset \fna^{-1}V_i$ then $\psi=\sum \psi_i$ and
$\fna^\star\psi=\phi$. 
\end{proof}

\begin{lemma}
\label{lm_supp}
Let $\fna\colon N\hookrightarrow M$ be an embedding with $\fna(N)\subset M$ closed
then
\begin{equation}
\supp\big(\fna_\varsigma(\DD\alpha)\big)=\fna\big(\supp(\alpha)\big)
\label{Push_support}
\end{equation}
\end{lemma}
\begin{proof}
Lemma \ref{lm_Dist_reg_supp} implies
$\supp(\DD\alpha)=\supp(\alpha)$ so 
$\fna\big(\supp(\DD\alpha)\big)=\fna\big(\supp(\alpha)\big)$.
Lemma \ref{lm_Push_supp} implies
$\supp\big(\fna_\varsigma(\DD\alpha)\big)\subset
\fna\big(\supp(\DD\alpha)\big)$. Thus 
$\supp\big(\fna_\varsigma(\DD\alpha)\big)\subset\fna\big(\supp(\alpha)\big)$.

By contrast let $x=\fna(y)\in \fna(\supp(\alpha))$ so
$y\in\supp(\alpha)$ as $\fna$ is injective. Given any neighbourhood
$U\subset M$ of $x$ then $\fna^{-1}(U)\subset N$ is a neighbourhood of
$y$. As $y\in\supp(\alpha)$ there exists $\psi\in\Gamma_0\Lambda N$ with
$\supp(\phi)\subset \fna^{-1} U$ such that $\int_N \phi\wedge\alpha\ne
0$. From lemma \ref{lm_embed_fna_star_inv} there exists a
$\psi\in\Gamma_0\Lambda M$ such that $\supp(\psi)\subset U$ and
$\fna^\star\psi=\phi$. Thus
$\fna_\varsigma(\DD\alpha)[\psi]=\DD\alpha[\fna^\star\psi]=
\DD\alpha[\phi]=\int_N
\phi\wedge\alpha\ne0$. Since this is true for all neighbourhoods $U$
about $x$ then $x\in\supp(\fna_\varsigma(\DD\alpha))$. Thus we have
shown $x\in\fna\big(\supp(\alpha)\big)\Longrightarrow 
x\in\supp\big(\fna_\varsigma(\DD\alpha)\big)$ hence
(\ref{Push_support}).
\end{proof}

\begin{lemma}
\label{lm_embed_alphaD}
If $\fna\colon N\hookrightarrow M$ is an embedding, 
$\fna(N)\subset M$ is closed and $\alpha\in\Gamma\Lambda N$ then 
$\fna_\varsigma(\DD\alpha)=0$ if and only if $\alpha=0$. 
\end{lemma}
\begin{proof}
Follows trivially from lemma \ref{lm_supp}.
\end{proof}

A \defn{submanifold distribution} is a distribution of the form 
$\Psi=\fna_\varsigma(\DD\alpha)$ where
\begin{eqnarray}
\fl\qquad\qquad
&\bullet\quad\textup{$\fna\colon N\hookrightarrow M$ is an embedding.}
\label{SMD_def_embded}
\\\fl
&\bullet\quad\textup{
$\alpha\in\Gamma\Lambda N$ with $\supp(\alpha)=N$.}
\label{SMD_def_supp}
\\\fl
&\bullet\quad\textup{
$\fna(N)\subset M$ is closed.}
\label{SMD_def_closed}
\\\fl
&\bullet\quad\textup{
$\fna\colon N\to \fna(N)$ is a diffeomorphism.}
\label{SMD_def_diffeo}
\end{eqnarray}
The set of all submanifold distributions over $M$ is written
$\GammaSD\Lambda M$. 

Since the combination of a pushforward of a regular distribution, for
example $\fna_\varsigma(\DD(\alpha))$ is so common, we introduce the
notation (the bold subscript $\varsigmaD$) to represent the
pushforward of a regular distribution so that
$\fna_\varsigmaD\alpha=\fna_\varsigma(\DD(\alpha))$.

Observe that if $\fna\colon N\hookrightarrow M$ and 
$\fna_\varsigmaD\alpha\in\GammaSD\Lambda M$ is a submanifold
distribution then from lemma \ref{lm_SMD_fna_proper} $\fna$ is proper and
from lemma \ref{lm_supp}
\begin{equation}
\supp(\fna_\varsigmaD\alpha)=\fna(N) 
\label{Push_SMD_supp}
\end{equation}

Given  $\Psi\in\GammaSD\Lambda M$,
the following lemma establishes the essential uniqueness of the
embedding and the form on the domain of the embedding.
\begin{lemma}
Given two submanifold distributions
$\fna_\varsigmaD\alpha\in\GammaSD\Lambda M$ and 
$\fnb_\varsigmaD\beta\in\GammaSD\Lambda M$ with $\fna\colon N\hookrightarrow M$
and $\fnb\colon P\hookrightarrow M$ then $\fna_\varsigmaD\alpha=\fnb_\varsigmaD\beta$
if and only if there exist a diffeomorphism $\fnc\colon N\to P$ with
$\alpha=\fnc^\star\beta$ and the following diagram commutes
\begin{equation*}
\xymatrix{
N \ar[rr]^\sfnc\ar[rd]_\sfna && P\ar[ld]^\sfnb \\ &M&}
\end{equation*}

\end{lemma}
\begin{proof}
Since $\supp(\fna_\varsigmaD\alpha)=\supp(\fnb_\varsigmaD\beta)$ then
$\fna(N)=\fnb(P)$. Furthermore $\fna\colon N\to \fna(N)$ and $\fnb\colon P\to
\fnb(P)$ are diffeomorphisms. Thus we can let
$\fnc=(\fnb|_{\fnb(P)})^{-1}\circ\fna\colon N\to P$, so $\fnc$ is a
diffeomorphism and $\fna=\fnb\circ \fnc$.

Given $\phi\in\Gamma_0\Lambda P$, from lemma \ref{lm_embed_fna_star_inv}
there exists $\psi\in\Gamma_0\Lambda M$ such that
$\fnb^\star\psi=\phi$. Now
\begin{eqnarray*}
\qquad\fl\int_P \phi\wedge \beta 
&=
\int_P \fnb^\star\psi\wedge \beta 
=
\fnb_\varsigmaD(\beta)[\psi]
=
\fna_\varsigmaD(\alpha)[\psi]
=
\int_N \fna^\star\psi\wedge\alpha 
=
\int_N (\fnb\circ \fnc)^\star\psi\wedge\alpha
\\\fl&=
\int_N (\fnc^\star \fnb^\star \psi)\wedge\alpha
=
\int_N \fnc^\star \phi\wedge\alpha
=
\int_P \fnc^{-1\star}( \fnc^\star \phi\wedge\alpha)
=
\int_P \phi\wedge (\fnc^{-1\star}\alpha)
\end{eqnarray*}
Since this is true for all $\phi$ then $\beta=\fnc^{-1\star}\alpha$.
\end{proof}

The following lemmas relate embeddings with internal contraction.

\begin{lemma}
\label{lm_v_transv_ne0}
Let $\fna\colon N\hookrightarrow M$ and
$\fna_\varsigmaD\alpha\in\GammaSD\Lambda M$. Let $x\in \fna(N)$ and
let $v\in\Gamma T M$ such that $v|_x\not\in T_x(\fna(N))$, that is $v$
is transverse to $\fna$ at $x$. Then $i_v (\fna_\varsigmaD\alpha)\ne
0$.
\end{lemma}
\begin{proof}
There exists an open neighbourhood $U\subset M$ of $x$ such that $v|_y$ is
transverse to $\fna_\varsigmaD\alpha$ for all $y\in \fna(N)\inter U$.
Shrinking $U$ if necessary we can assume that 
$U\inter\fna(\partial N)=\emptyset$.
Further shrinking $U$ we can make it contain a coordinate chart
about $x$ adapted to $v$. Thus there exists a $t\colon U\to \Real$ such that
$t(y)=0$ for all $y\in \fna(N)$ and $v\Vacts(t)=1$.

Now for $\psi\in\Gamma_0\Lambda M$ with $\supp(\psi)\subset U$ then
\begin{eqnarray*}
\fna_\varsigmaD\alpha[dt\wedge\psi]
&=
\int_N \fna^\star(d t\wedge\psi)\wedge\alpha
=
\int_N d \fna^\star(t)\wedge\fna^\star(\psi)\wedge\alpha
\\
&=
\int_{\partial N} \fna^\star(t)\wedge\fna^\star(\psi)\wedge\alpha
-\int_N \fna^\star(t)\, d(\fna^\star(\psi)\wedge\alpha)
=
0
\end{eqnarray*}
since $\fna^\star t(z)=t(\fna(z))=0$ as $\fna(z)\in\fna(N)$, and 
since also $\fna^\star\psi|_{\partial N}=0$.

There exists
$\phi\in\Gamma_0\Lambda M$ such that $\supp(\phi)\subset U$ and
$\fna_\varsigmaD\alpha[\phi]\ne0$. To see this consider the contrary that
$\supp(\phi)\subset U$ implied $\fna_\varsigmaD\alpha[\phi]=0$. This would imply that 
$U\subset M\backslash\supp(\fna_\varsigmaD\alpha)$, a contradiction.

Now consider $i_v(\fna_\varsigmaD\alpha)[\phi\wedge dt]$
\begin{eqnarray*}
\fl\qquad
i_v(\fna_\varsigmaD\alpha)[\phi\wedge dt] &=
-i_v(\fna_\varsigmaD\alpha)[(dt\wedge\phi)^\eta] =
\fna_\varsigmaD\alpha[i_v(dt\wedge\phi)] =
\fna_\varsigmaD\alpha[\phi] -
\fna_\varsigmaD\alpha[dt\wedge i_v\phi] \\\fl&=
\fna_\varsigmaD\alpha[\phi] 
\ne 0
\end{eqnarray*}
Hence $i_v(\fna_\varsigmaD\alpha)\ne 0$.

\end{proof}

\begin{lemma}
\label{lm_intg_surf}
Let $\fna_\varsigmaD(\alpha)\in\GammaSD\Lambda M$ with 
$\fna\colon N\hookrightarrow M$.
Then
\begin{equation}
i_v \fna_\varsigmaD\alpha = 0
\label{Push_i_v_0}
\end{equation}
if and only if $v$ is tangential to $\fna$ and
\begin{equation}
i_u\alpha=0
\label{Push_i_fv_0}
\end{equation}
where $u\in\Gamma TN$ is the unique vector field satisfying
$\fna_\star(u)=v|_{\fna(N)}$.
\end{lemma}
\begin{proof}
That (\ref{Push_i_fv_0}) implies (\ref{Push_i_v_0}) follows trivially
from lemma \ref{lm_Push_iv_comm}.

If (\ref{Push_i_v_0}) is true and $x\in \fna(N)$ 
then 
from lemma \ref{lm_v_transv_ne0} $v|_x$ must be tangential to
$\fna$. Then lemma \ref{lm_Push_iv_comm} gives
$\fna_\varsigmaD(i_u\alpha)=0$. Finally (\ref{Push_i_fv_0}) follows from
lemma \ref{lm_embed_alphaD}.
\end{proof}

\vspace{0.5em}

The set of $p-$form distributions $\GammaD\Lambda^p M$ forms a vector
space in that it is closed under addition and multiplication by a
scalar. Also the exterior derivative $d\colon \GammaD\Lambda^{p}M\to
\GammaD\Lambda^{p+1}M$. By contrast the set of $p-$form submanifold
distributions $\GammaSD\Lambda^p M$ does not in general form a vector
space. For example if $\fna_\varsigmaD\alpha\in\GammaSD\Lambda^p M$
and $\fnb_\varsigmaD\beta\in\GammaSD\Lambda^p M$ with $\fna\colon
N\hookrightarrow M$, $\fnb\colon P\hookrightarrow M$ and $\dim
N \ne \dim P$ then
$\fna_\varsigmaD\alpha+\fnb_\varsigmaD\beta\notin\GammaSD\Lambda^p M$.
Also if $\dim N=\dim P$ and $\fna(N)\inter \fnb(P)\ne\emptyset$ then
in general
$\fna_\varsigmaD\alpha+\fnb_\varsigmaD\beta\notin\GammaSD\Lambda^p M$.

Likewise in section
\ref{ch_Bdd} we see that the exterior derivative
does not map $\GammaSD\Lambda^{p}M$ to $\GammaSD\Lambda^{p+1}M$.
This is a due to the possible additional boundary terms. 


\subsection{Pullback of Distributions}
\label{ch_Pull}

We first define the pullback for a general ``pushforward''
distribution. This is a distribution of the form
$\fna_\varsigmaD\alpha\in\GammaD\Lambda M$. This is required for the
proof of lemma \ref{lm_fold_push}. However in most cases the map
$\fna$ is an embedding, and this simplifies the concept. 
This definition of a pullback of a distribution can be 
compared to the definition using the weak limit \cite{Friedlander}.

Let $\fna\colon N\to M$ and $\fnb\colon P\to M$ be proper maps and let
$\alpha\in\Gamma\Lambda N$. Thus 
$\fna_\varsigmaD\alpha\in\GammaD\Lambda M$ is a distribution and we
wish to define the pullback 
$\fnb^\varsigma(\fna_\varsigmaD\alpha)\in\GammaD\Lambda P$ with respect to
$\fnb$. Let
\begin{equation}
Q=\Set{(y,p)\in N\times P\big| \fna(y)=\fnb(p)}
\label{AltPull_def_Q}
\end{equation}
be the induced manifold (also know as the pullback manifold),
such that the following diagram commutes
\begin{equation}
\xymatrix{
Q \ar[r]^{\hat \sfnb} \ar[d]_{\hat \sfna}
& N \ar[d]^{\sfna}
\\
P \ar[r]^\sfnb & M
}
\qquad
\textup{\raisebox{-2em}{where $\hat \fnb(y,p)=y$ and $\hat \fna(y,p)=p$.}}
\label{Pull_comm_diag}
\end{equation}
We say that
$\fna_\varsigmaD\alpha\in\GammaD\Lambda M$ is \defn{transverse} to $\fnb$
if
\begin{equation}
\dim Q =
\dim N+\dim P - \dim M
\label{AltPull_dim}
\end{equation}
Given $\fna_\varsigmaD\alpha\in\GammaSD\Lambda M$ transverse to $\fnb$
then we can define the \defn{pullback} of $\fna_\varsigmaD\alpha$ by
$\fnb$
\begin{equation}
\fnb^\varsigma\big(\fna_\varsigmaD\alpha\big)=
\hat \fna_{\varsigmaD}(\hat \fnb^\star\alpha)\in\GammaD\Lambda P
\label{AltPull_def__fna_pull}
\end{equation}

\begin{lemma}
The pullback preserves the degree.
\end{lemma}
\begin{proof}
From (\ref{Push_deg}) we have
\begin{eqnarray*}
\fl\qquad
\deg\big(\fnb^\varsigma(\fna_\varsigmaD\alpha)\big)
&=
\deg\big(\hat \fna_{\varsigmaD}(\hat \fnb^\star\alpha)\big)
=
\deg(\hat\fnb^\star \alpha) + \dim P -\dim Q
\\\fl&=
\deg(\alpha) + \dim M - \dim N
=
\deg\big(\fna_\varsigmaD\alpha\big)
\end{eqnarray*}
\end{proof}

\begin{lemma}
For regular distributions we observe
\begin{equation}
\fna^\varsigma(\DD\alpha)=\DD(\fna^\star\alpha)
\label{Pull_regular}
\end{equation}
\end{lemma}
\begin{proof}
For regular distributions (\ref{Pull_comm_diag}) becomes
\begin{equation*}
\xymatrix{
P \ar[r]^{\sfnb} \ar[d]_{\sId}
& M \ar[d]^{\sId}
\\
P \ar[r]^\sfnb & M
}
\end{equation*}
Hence for $\alpha\in\Gamma\Lambda M$,
$\fna^\varsigma(\DD\alpha)=\Id_\varsigmaD(\fna^\star\alpha)
=\DD(\fna^\star\alpha)$.
\end{proof}

\begin{lemma}
\label{lm_Pull_comp}
Composition of pullbacks: Let $\fnb\colon P\to M$ and 
$\fnc\colon Q\to P$, where $\fnb^\varsigma$ and $\fnc^\varsigma$ are
defined, then
\begin{equation}
(\fnb\circ\fnc)^\varsigma=\fnc^\varsigma\circ\fnb^\varsigma
\label{Pull_comp}
\end{equation}
\end{lemma}
\begin{proof}
Let $\fna_\varsigmaD\alpha\in\GammaD\Lambda M$ with 
$\fna\colon N\to M$ and $\alpha\in\Gamma\Lambda N$. Define
$\hat P,\hat Q,\hat \fna,\hat \fnb,\hat \fnc,\tilde \fna$ using
(\ref{AltPull_def_Q},\ref{Pull_comm_diag}) so that the following
commutes
\begin{equation*}
\xymatrix{
\hat Q \ar[r]^{\hat \sfnc} \ar[d]_{\tilde\sfna}
&
\hat P \ar[r]^{\hat \sfnb} \ar[d]_{\hat \sfna}
& N \ar[d]^{\sfna}
\\
Q \ar[r]^\sfnc &
P \ar[r]^\sfnb & M
}
\end{equation*}
then $\dim \hat Q -\dim Q = \dim \hat P-\dim P = \dim N - \dim M$, so
that the pullbacks are defined.
\begin{equation*}
(\fnb\circ\fnc)^\varsigma (\fna_\varsigmaD\alpha)
=
\tilde\fna_\varsigmaD \big( (\hat\fnb\circ\hat\fnc)^\star \alpha\big)
=
\tilde\fna_\varsigmaD \big(\hat\fnc^\star(\hat\fnb{}^\star \alpha)\big)
=
\fnc^\varsigmaD\big(\fna_\varsigmaD (\hat\fnb{}^\star \alpha)\big)
=
\fnc^\varsigmaD\big(\fnb^\varsigmaD (\fna_\varsigmaD \alpha)\big)
\end{equation*}
hence (\ref{Pull_comp}).

\end{proof}

It is often the case that both $\fna\colon N\hookrightarrow M$ and 
$\fnb\colon P\hookrightarrow M$ are embeddings. 
Observe that in this case
the intersection $Q\cong \fna(N)\inter \fnb(P)$ since
\begin{equation*}
\fna(\hat \fnb(Q))=\fnb(\hat \fna(Q))=\fna(N)\inter \fnb(P)
\end{equation*}
and all these maps are embeddings.

\subsection{Boundaries}
\label{ch_Bdd}

\begin{lemma}
\label{lm_Bdd}
Let $N$ be a manifold with $\dim N=n$ and have boundary $B=\partial N$
with $\iota\colon B\to N$ and let $\alpha\in\Gamma\Lambda^p N$
($\alpha$ is smooth). Then
\begin{equation}
d(\DD\alpha) 
= 
\DD(d\alpha) + (-1)^{n-p} \iota_\varsigmaD(\iota^\star\alpha)
\label{Push_d_Nbdd}
\end{equation}
\end{lemma}
\begin{proof}
Let $\phi\in\Gamma_0\Lambda^{n-p-1} N$ then
\begin{eqnarray*}
\fl\qquad
d(\DD\alpha)[\phi]
&=
-\DD\alpha[d\phi^\eta]
=
-\int_N d\phi^\eta\wedge\alpha
=
-\int_N d(\phi^\eta\wedge\alpha)
+
\int_N \phi\wedge d\alpha
\\
&=
-\int_B \iota^\star(\phi^\eta\wedge\alpha)
+
\DD(d\alpha)[\phi]
=
\DD(d\alpha)[\phi]
+
(-1)^{n-p}
\int_B \iota^\star(\phi)\wedge\iota^\star(\alpha)
\\
&=
\DD(d\alpha)[\phi]
+
(-1)^{n-p}
\DD(\iota^\star\alpha)[\iota^\star(\phi)]
\\&=
\DD(d\alpha)[\phi]
+
(-1)^{n-p}
\iota_\varsigmaD(\iota^\star\alpha)[\phi]
\end{eqnarray*}
\end{proof}

We observe that in general the right hand side of (\ref{Push_d_Nbdd})
is not a submanifold distribution, since the domain of $D(d\alpha)$ is
$N$ whereas the domain of $\iota_\varsigmaD(\iota^\star\alpha)$ is
$B$ and $\dim N\ne\dim B$.

Also in general $\iota_\varsigmaD(\iota^\star\alpha)$ is not a
submanifold distribution. This is because in general
$\iota\colon B\to N$ is not an embedding. For example, if $N$ is
a 3-dimensional solid bounded cylinder then $B$ consists of two discs
and a cylinder $S^1\times I$ where $I$ is a closed interval, and
$\iota\colon B\to N$ is not injective.

Although this doesn't matter for smooth forms since the set where
$\iota$ is not injective has measure zero, one has to be careful when
dealing with distributional forms.

\begin{lemma}
\label{lm_Bdd_dalpha}
Let $\alpha\in\Gamma\Lambda N$, so that $\alpha$ is smooth then
\begin{equation}
d(\DD\alpha)=0 \quad\Longrightarrow\quad
d\alpha=0
\label{Bdd_dD}
\end{equation}
\end{lemma}
\begin{proof}
Let $B=\partial N$ with $\iota\colon B\to N$, 
$\deg(\alpha)=p$ and $\dim N=n$. Given $\phi\in\Gamma_0\Lambda M$ with
$\supp(\phi)\subset N\backslash B$ then 
\begin{eqnarray*}
\fl\qquad
0&=
d(\DD\alpha)[\phi]=
\DD(d\alpha)[\phi]+(-1)^{n-p}\iota_\varsigmaD(\iota^\star\alpha)[\phi]
=
\DD(d\alpha)[\phi]+(-1)^{n-p}\DD(\iota^\star\alpha)
[\iota^\star\phi]
\\\fl&=
\DD(d\alpha)[\phi]
\end{eqnarray*}
since $\iota^\star\phi=0$. Thus $\supp\big(\DD(d\alpha)\big)\subset
B$. From lemma \ref{lm_Dist_reg_supp}, $\supp (d\alpha)\subset
B$. So $d\alpha=0$ on $N\backslash B$, but since $d\alpha$ is continuous and
$B$ contains no open sets, $d\alpha=0$.
\end{proof}

If $\fna\colon N\to M$ and neither $M$ nor $N$ has a boundary then
we have the trivial result following from (\ref{Dist_D_basic_res})
lemma \ref{lm_Push_d_comm}
\begin{equation}
d (\fna_\varsigmaD\alpha)=\fna_\varsigmaD(d\alpha)
\label{Bdd_no_bdd_d}
\end{equation}

\section{Distributional solutions to transport equations}
\label{ch_Trans}

Given a manifold $M$, with $\dim M=n$ 
and a nowhere zero vector field $v\in\Gamma TM$ then we say
that the distribution $\Psi\in\GammaD\Lambda^{n-1}M$ is a
solution to the transport equations if 
\begin{equation}
d\Psi=0\qquadand
i_v\Psi=0
\label{Trans_transp_eqns}
\end{equation}
We use the name transport equations to refer to {\em any} vector
field $v\in\Gamma TM$.  The transport equations with respect to the
vector field $\LiouV\in\Gamma T\Ebun$ where $\pi\colon\Ebun\to\Man$,
$\Ebun\subset TM$ is a bundle (usually the upper unit hyperboloid over
spacetime) and $\LiouV$ is horizontal, that is $\pi_\star(\LiouV|_u)=u$
for $u\in\Ebun$, are also known as the Liouville equation, the
collisionless Boltzmann equation and the Vlasov equation
(\ref{MV_pdd_closed}-\ref{MV_Liou_eqn}).

In general we will only be interested in solutions to the transport
equations which are submanifold distributions and which, if they
possess a boundary then the boundary is an embedding. See figure
\ref{fig_soln}. These are given by
\begin{theorem}
\label{lm_Trans_soln}
Let $N$ be a manifold which is either without boundary or if it 
possess a boundary $B$ then let $\iota\colon B\hookrightarrow N$ be a closed
embedding and let $M$ be without boundary. Let $\fna\colon N\hookrightarrow M$ 
and $\alpha\in\Gamma\Lambda^{\dim N-1} M$.
Then the submanifold distribution
$\Psi=\fna_\varsigmaD\alpha\in\GammaSD\Lambda M$
satisfies the transport equations (\ref{Trans_transp_eqns}) if and only
if the following three conditions hold
\begin{eqnarray}
\fl\qquad\qquad
\bullet &\quad
d\alpha=0
\label{Trans_soln_cond_d_alpha}
\\
\fl\qquad\qquad
\bullet &\quad
\textup{There exists $u\in\Gamma T N$ such that 
$\fna_\star(u)=v|_{\fna(N)}$ and
$i_u\alpha=0$.}
\label{Trans_soln_cond_iu}
\\
\fl\qquad\qquad
\bullet &\quad
\textup{If $N$ has a boundary  then
$u$ is tangential to $\iota$.}
\label{Trans_soln_cond_bdd}
\end{eqnarray}
\end{theorem}
\begin{proof}
First show that the three conditions
(\ref{Trans_soln_cond_d_alpha}-\ref{Trans_soln_cond_bdd}) imply
that $\Psi$ is a solution to the transport equations
(\ref{Trans_transp_eqns}).
Let $w\in\Gamma T B$ satisfy $\iota_\star w = u|_{\iota(B)}$ then
\begin{equation*}
i_w\iota^\star\alpha=
\iota^\star (i_{\iota_\star w}\alpha)=
\iota^\star (i_u\alpha)=
\iota^\star 0=
0
\end{equation*}
Now since $\deg(\iota^\star\alpha)=\dim N-1=\dim(B)$ and $w\ne 0$ then
$\iota^\star\alpha=0$.

From lemmas \ref{lm_Push_d_comm} and \ref{lm_Bdd}
\begin{equation*}
d\Psi=d \fna_\varsigmaD\alpha=\fna_\varsigma (d(\DD\alpha))=
\fna_\varsigmaD(d\alpha)-
\fna_\varsigma \iota_\varsigmaD (\iota^\star\alpha)
=
0
\end{equation*}
and from lemma \ref{lm_Push_iv_comm}
\begin{equation*}
i_v\Psi=i_v \fna_\varsigmaD\alpha=\fna_\varsigma(i_u (\DD\alpha))=
\fna_\varsigma(\DD(i_u\alpha))=0
\end{equation*}

In order to show that the transport equations
(\ref{Trans_transp_eqns}) imply 
(\ref{Trans_soln_cond_d_alpha}-\ref{Trans_soln_cond_bdd}):
From  (\ref{Trans_transp_eqns}) we have
$0=i_v\Psi=i_v \fna_\varsigmaD\alpha$.
Hence from lemma \ref{lm_intg_surf}, $v$ is tangential to $\fna$ and we have
(\ref{Trans_soln_cond_iu}). 

Let $\iota\colon B\hookrightarrow N$ be the boundary of $N$. Then 
$0=d(\fna_\varsigmaD\alpha)=\fna_\varsigma(d(\DD\alpha))$. From lemma
\ref{lm_embed_alphaD} $0=d(\DD\alpha))$ hence from lemma
\ref{lm_Bdd_dalpha} $d\alpha=0$ and hence
$\fna_\varsigma\iota_\varsigmaD(\iota^\star\alpha)=0$.
Since $\fna\circ\iota\colon B\hookrightarrow M$ is a closed embedding and
$0=\fna_\varsigma \iota_\varsigmaD(\iota^\star\alpha)=
(\fna\circ\iota)_\varsigmaD(\iota^\star\alpha)$, it follows from lemma
\ref{lm_embed_alphaD} that $\iota^\star\alpha=0$. 

Now $\alpha$ satisfies the transport equations with respect to $u$,
i.e. $d\alpha=0$ and $i_u\alpha=0$.  If $u$ is transverse to $\iota$
at any point in $B$ then $u$ is transverse to $\iota$ in an open
subset $V\subset B$. Furthermore since $\iota^\star\alpha=0$,
$i_u\alpha=0$ and $u$ is transverse to $\iota$ then $\alpha|_y=0$ for
all $y\in V$. Using the transport equations for $\alpha$ this implies
there exists a an open subset $U\subset N$ such that
$\alpha|_{U}=0$. This contradicts the statement that
$\supp(\alpha)=N$. Hence $u$ is tangential to $\iota$.
\end{proof}

An \defn{initial hypersurface} associated with $v$ is a submanifold
$\sigma\colon \Sigma\hookrightarrow M$ such that each integral curve
of $v$ intersects $\Sigma$ exactly once, and $v$ contains no closed
curves. For the following we further demand that each integral curve
of $v$ has domain $\Real$. If $\tau$ is the parameter along an
integral curve $\gamma_x\colon \Real\to M$, with
$\gamma_x(0)=x\in\Sigma$, then
$v|_{\gamma_x(\tau)}=\gamma_{x\star}(\partial_\tau|_{\tau})=\dot\gamma_x(\tau)$.
These two conditions imply that there is a diffeomorphism $\fnc\colon
\Real\times\Sigma\to M$ with $\fnc(\tau,x)=\gamma_x(\tau)$.

For $M=\Ebun$ and $v=\LiouV$, the Liouville vector field
(\ref{MV_def_W}), then since $\LiouV$ is horizontal the requirement
that the domain of $\gamma_x$ is $\Real$ is equivalent to demanding
that the proper time for each curve $C_x$ where $\Cd_x=\gamma_x$ is
from $-\infty$ to $+\infty$.

A \defn{box} $s\colon S\hookrightarrow M$ is any compact submanifold
(usually with boundary) of an initial hypersurface and of the same
dimension as the hypersurface. That is $v$ is transverse to $s$ and
$\dim S=\dim M-1$.  Given a box $s\colon S\hookrightarrow M$ we can
define the \defn{expectation of $\Psi$} as
\begin{equation}
[s^\varsigma(\Psi)]=
(s^\varsigma\Psi)[1]
\end{equation}
which is valid since $1\in\Gamma\Lambda S$ has compact support on $S$.

The \defn{future} of a box or an initial hypersurface is the set of
points which lie on the integral curves of $v$ in the direction of $v$
away from $S$ or $\Sigma$.

The total \defn{charge} on an initial hypersurface
$\sigma\colon \Sigma\hookrightarrow M$ associated with $\Psi$ is given by
\begin{equation}
Q_{\textups{total}}=[\sigma^\varsigma(\Psi)]
\label{Trans_def_charge}
\end{equation}
However in order to define charge we require that given a series of
boxes $S_i\subset S_{i+1}$ with $\Union S_i=\Sigma$ then
$\lim_{i\to\infty} [s_i^\varsigma(\Psi)]$ is well defined and
independent of the choice of $\Set{S_1,S_2,\ldots}$.

\begin{theorem}
\label{lm_Trans_S12}
Given a solution $\fna_\varsigmaD\alpha\in\GammaSD\Lambda M$ to the
transport equations where $\fna\colon N\hookrightarrow M$ is an
embedding. Let $s_1\colon S_1\hookrightarrow M$ and $s_2\colon
S_2\hookrightarrow M$ be two boxes with the property that any integral
curve of $v$ passing through one also passes through the other.  Then
\begin{equation}
[s_1^\varsigma (\fna_\varsigmaD\alpha)]=
[s_2^\varsigma (\fna_\varsigmaD\alpha)]
\label{Trans_S12}
\end{equation}
\end{theorem}
\begin{proof}
Assume first that $S_2$ lies to the future of $S_1$.  Let
$\iota\colon U\hookrightarrow M$ be the closed submanifold with $\dim U=\dim
M$, given by the union of all the integral curves of $v$ between $S_1$
and $S_2$. The pullback manifold with respect to $\iota$ and $\fna$ is
given by $\fna^{-1}(U)$ and (\ref{Pull_comm_diag}) becomes (defining
the embeddings $\hat\iota$, $\hat \fna$, $\hat \fna_i$ and $\hat
s_i$.)
\begin{equation*}
\xymatrix{
\fna^{-1}(S_i) \arhook[r]^{} \arhook[d]^{\hat{\sfna}_i}
\arhook@/^1.5pc/[rr]^{\hat s_i}
&
\fna^{-1}(U) \arhook[r]_{\hat\iota} \arhook[d]^{\hat \sfna}
& 
N \arhook[d]^{\sfna}
\\
S_i \arhook[r]^{} \arhook@/_1.8pc/[rr]_{s_i}
&
U \arhook[r]^{\iota} & M
}  
\end{equation*}
By definition $s_i^\varsigma
(\fna_\varsigmaD\alpha)=(\hat{\fna}_i)_\varsigmaD(\hat{s_i}^\star\alpha)$ so
\begin{equation*}
[s_i^\varsigma (\fna_\varsigmaD\alpha)]
=
s_i^\varsigma (\fna_\varsigmaD\alpha)[1]
=
(\hat{\fna}_i)_\varsigmaD(\hat{s_i}^\star\alpha)[1]
=
D(\hat{s_i}^\star\alpha)[1]
=
\int_{\fna^{-1}(S_i)} \hat{s_i}^\star\alpha
\end{equation*}

Now $\alpha\in\Gamma\Lambda N$ so
$\hat\iota^\star\alpha\in\Gamma\Lambda (\fna^{-1}U)$ is a regular
form. Therefore we can perform the standard analysis.  From the
transport equations then (\ref{Trans_soln_cond_d_alpha}) implies
$d\alpha=0$ and (\ref{Trans_soln_cond_iu}) implies $i_u\alpha=0$ where
$u\in\Gamma T N$ is the unique vector field satisfying
$\fna_\star(u)=v|_{\fna(N)}$.

The boundary of $U$ is given by
\begin{equation*}
\partial(U)=S_2-S_1+V
\end{equation*}
where the $-$ sign refers to the orientation of $S_1$ and $u$ is
tangential to $V$. Since $\fna$ is an embedding
the boundary of $\fna^{-1}(U)$ is given by
\begin{equation*}
\partial(\fna^{-1}(U))=\fna^{-1}(S_2)-\fna^{-1}(S_1)+\fna^{-1}(V)
\end{equation*}
where $u$ is tangential to $\fna^{-1}(V)$. Therefore $\int_{\fna^{-1}(V)}
\alpha=0$. Thus
\begin{equation*}
0=\int_{\fna^{-1}U} d\alpha
= \int_{\fna^{-1}(S_2)} \alpha-\int_{\fna^{-1}(S_1)} \alpha+\int_{\fna^{-1}(V)} \alpha
= [s_1^\varsigma (\fna_\varsigmaD\alpha)]-
[s_2^\varsigma (\fna_\varsigmaD\alpha)]
\end{equation*}

For two general boxes $S_1$ and $S_2$ simply take another box $S_3$
lying to the future of both $\Sigma_1$ and $\Sigma_2$.
\end{proof}

\begin{corrol}
Conservation of charge. Given two initial hypersurfaces
$\sigma\colon \Sigma\hookrightarrow M$ and
$\hat\sigma\colon \hat\Sigma\hookrightarrow M$ and given that the charge
(\ref{Trans_def_charge}) is well defined with respect to $\Sigma$
then the charge is well defined with respect to $\hat\Sigma$ and
\begin{equation}
[\hat\sigma^\varsigma(\Psi)]=[\sigma^\varsigma(\Psi)]
\label{Trans_charge_conserve}
\end{equation}
\end{corrol}
\begin{proof}
Any box $s\colon S\hookrightarrow\Sigma$ corresponds to a box 
$\hat s\colon \hat S\hookrightarrow \hat\Sigma$. 
Therefore from theorem \ref{lm_Trans_S12}
$[s^\varsigma(\Psi)]=[\hat s^\varsigma(\Psi)]$. Taking the limit of
boxes shows that $[\sigma^\varsigma(\Psi)]$ is well defined and that
(\ref{Trans_charge_conserve}) holds.

\end{proof}

Given a submanifold distribution on an initial hypersurface we can generate
a unique solution to the transport equations.

\begin{theorem}
\label{thm_Transp_initial_distr}
Let $\sigma\colon \Sigma\hookrightarrow M$ be an initial hypersurface of
$v\in\Gamma TM$ and let $\fnc\colon \Real\times\Sigma\to M$ be the
corresponding diffeomorphism.
Let $\fnb_\varsigmaD\alpha\in\GammaSD\Lambda\Sigma$ be
a submanifold distribution, with $\fnb\colon N\hookrightarrow\Sigma$
and $\alpha\in\Gamma\Lambda^{\dim N}N$.  
Let
\begin{equation}
\Psi = \fna_\varsigmaD(\pr_2^\star\alpha)\in\GammaSD\Lambda^{\dim M-1}M
\label{Trans_soln_pushed} 
\end{equation}
where
\begin{equation}
\fna\colon \Real\times N\to M\;;\qquad \fna(\tau,y)=\fnc(\tau,\fnb(y))
\label{Trans_soln_def_S}
\end{equation}
and $\pi_2\colon \Real\times N\to N$ is the natural projection.
Then $\Psi$ satisfies the transport equations
(\ref{Trans_transp_eqns}) and the initial conditions 
$\sigma^\varsigma\Psi=\fnb_\varsigmaD\alpha$.

Furthermore $\Psi$ is unique in that given any $\Phi\in\GammaSD\Lambda
M$ which also satisfies the transport equations, and
$\sigma^\varsigma\Phi=\fnb_\varsigmaD\alpha$, then $\Psi=\Phi$.
\end{theorem}
\begin{proof}
To establish that $\Psi$ is indeed a submanifold distribution, we must
show the four conditions
(\ref{SMD_def_embded}-\ref{SMD_def_diffeo}). The following
commutes
\begin{equation*}
\xymatrix{
\Real\times N \ar[rd]^\sfna \ar[d]_{\sId\times \sfnb} &
\\
\Real\times\Sigma \ar[r]_\sfnc & M
}
\end{equation*}
Thus $\fna$ is an embedding.
Since $\Real\times \fnb(N)\subset\Real\times\Sigma$ is a
closed submanifold. Then $\fna(N\times\Real)\subset M$ is closed. Clearly
$\supp(\pi_2^\star\alpha)=\Real\times N$. Also since $\Id\times
\fnb\colon \Real\times N\to \Real\times \fnb(N)$ is a diffeomorphism so is $\fna$.

\vspace{1em}

To establish that $d\Psi=0$ it is necessary to realise that
$\Real\times N$ may have a boundary.
Let $\iota\colon B\hookrightarrow N$ be the boundary of $N$. Then the
boundary of $\Real\times N$ is given by 
$\hat\iota\colon \Real\times B\to \Real\times N$ where 
$\hat\iota=\Id\times\iota$,
i.e. $\hat\iota(\tau,y)=(\tau,\iota(y))$.
\begin{equation*}
d\Psi 
=
d \fna_\varsigmaD(\pr_2^\star\alpha)
=
d \fna_\varsigma\DD(\pr_2^\star\alpha)
=
\fna_\varsigma d\DD(\pr_2^\star\alpha)
=
\fna_\varsigmaD(d\pr_2^\star\alpha)
-
\fna_\varsigma\hat\iota_\varsigmaD(\hat\iota^\star\pi_2^\star\alpha)
\end{equation*}
Now $d\pr_2^\star\alpha=\pr_2^\star d\alpha=0$ since
$\deg(\alpha)=\dim N$. 

Let 
$\hat\pr_2\colon \Real\times B\to B$ be the second projection so that
$\pr_2\circ\hat\iota=\iota\circ\hat\pr_2$. Then
\begin{equation*}
\hat\iota^\star\pi_2^\star\alpha = 
(\pi_2\circ\hat\iota)^\star\alpha =
(\iota\circ\hat\pr_2)^\star\alpha =
\hat\pr_2^\star(\iota^\star\alpha)=0
\end{equation*}
since $\deg\alpha=\dim N>\dim B$. Thus $d\Psi=0$.

We now establish that $i_v\Psi=0$.
\begin{equation*}
i_v \fna_\varsigmaD (\pr_2^\star\alpha)
=
i_v \fna_\varsigma\DD (\pr_2^\star\alpha)
=
\fna_\varsigma i_{\partial_\tau}\DD(\pr_2^\star\alpha)
=
\fna_\varsigmaD (i_{\partial_\tau}\pr_2^\star\alpha)
=
\fna_\varsigmaD (\pr_2^\star i_{\pr_{2\star}\partial_\tau}\alpha)
=
0
\end{equation*}
since $\fna_\star(\partial_\tau)=v$ and $\pr_{2\star}\partial_\tau=0$.

\vspace{1em}

We now establish that $\Psi$ satisfies the initial conditions.
Let $I_0\colon N\hookrightarrow \Real\times N$, $I_0(y)=(0,y)$ then
the following commutes and is the pullback manifold
\begin{equation*}
\xymatrix{ 
N \arhook[r]^{I_0} \arhook[d]_{\sfnb} & 
\Real\times N  \arhook[d]^{\sfna} \\
\Sigma \arhook[r]_{\sigma} & M}  
\end{equation*}
and $I_0^\star(\pi_2^\star(\alpha))=\alpha$ so
$\sigma^\varsigma(\Psi)=\fnb_\varsigmaD\alpha$.

\vspace{1em}

We now establish that $\Psi$ is unique. We first show that
$\supp(\Psi)=\supp(\Phi)$. 
Let  $s\colon S\hookrightarrow M$ be a box so that $x\in S$ and
$S\inter\Sigma=\emptyset$. Now there lies a box 
$\hat s\colon \hat S\hookrightarrow\Sigma$
which lies to the future or past of $S$ hence from lemmas
\ref{lm_Trans_S12} and \ref{lm_Pull_comp} we have
\begin{equation*}
[s^\varsigma(\Phi)]=
[(\sigma\circ\hat s)^\varsigma(\Phi)]=
[\hat s^\varsigma(\sigma^\varsigma\Phi)]=
[\hat s^\varsigma(\sigma^\varsigma\Psi)]=
[(\sigma\circ\hat s)^\varsigma(\Psi)]=
[s^\varsigma(\Psi)]
\end{equation*}
Thus given $x\in\supp(\Phi)$, then for any box $s\colon S\hookrightarrow M$
with $x\in s(S)$ then $[s^\varsigma(\Psi)]=[s^\varsigma(\Phi)]\ne0$.
Thus $x\in\supp(\Psi)$ and visa versa.

Since $\supp(\Psi)=\supp(\Phi)$ we may consider up to diffeomorphism
that $\Phi=\fna_\varsigmaD\beta$ for some
$\beta\in\Gamma\Lambda(\Real\times N)$. From theorem
\ref{lm_Trans_soln}, both $\pi_2^\star\alpha$ and $\beta$ satisfy the
transport equations with respect to $\partial_\tau$ on $\Real\times
N$. Furthermore $\fnb_\varsigmaD(I_0^\star\pi_2^\star\alpha)=
\sigma^\varsigma\Psi=\sigma^\varsigma\Phi=
\fnb_\varsigmaD(I_0^\star\beta)$ hence 
$I_0^\star\beta=I_0^\star\pi_2^\star\alpha$, i.e. they agree
on the initial hypersurface $I_0\colon N\hookrightarrow
(\Real\times N)$. Hence $\beta=\pi_2^\star\alpha$ and so $\Phi=\Psi$.
\end{proof}

There are two alternative formulations of $\Psi$, one simply using the
pullback, a second using an integral (\ref{Trans_soln_pushed}) which
can easily be generalised for general distributions.
\begin{corrol}
\label{lm_Trans_pull_form}
Let $\sigma\colon \Sigma\hookrightarrow M$, $v\in\Gamma TM$,
$\fnc\colon \Real\times\Sigma\to M$, 
$\fnb_\varsigmaD\alpha\in\GammaSD\Lambda\Sigma$,
$\fnb\colon N\hookrightarrow\Sigma$,
$\alpha\in\Gamma\Lambda^{\dim N}N$ and   
$\Psi \in\GammaSD\Lambda^{\dim M-1}M$ be as in theorem
\ref{thm_Transp_initial_distr}. Then
\begin{equation}
\Psi=(\hat\pi_2\circ\fnc^{-1})^\varsigma(\fnb_\varsigmaD\alpha)
\label{Trans_pull_form}
\end{equation}
where $\hat\pi_2:\Real\times\Sigma\to\Sigma$ is the second projection.
\end{corrol}
\begin{proof}
Since $\fnc\colon \Real\times\Sigma\to M$ is a diffeomorphism then
$(\hat\pi_2\circ\fnc^{-1})\colon M\to\Sigma$. Thus to define the
pullback $(\hat\pi_2\circ\fnc^{-1})^\varsigma$ we have the following
commutative diagram.
\begin{equation*}
\xymatrix{ 
\Real\times N
\ar[r]^{\pi_2}
\arhook[d]_{\sfna}
&
N
\arhook[d]_{\sfnb}
\\
M
\ar[r]_{\hat\pi_2\circ\sfnc^{-1}}
& \Sigma
 }
\end{equation*}
Clearly $\dim (\Real\times N)+\dim\Sigma=(\dim N+1) + (\dim M-1)=\dim
N+\dim M$ so $\hat\pi_2\circ\fnc^{-1}$ is transverse to $\fnb$, and 
(\ref{Trans_pull_form}) follows from 
(\ref{Trans_soln_pushed}) and
(\ref{AltPull_def__fna_pull}).
\end{proof}

\begin{corrol}
\label{lm_Trans_int_soln}
Let $\sigma\colon \Sigma\hookrightarrow M$, $v\in\Gamma TM$,
$\fnc\colon \Real\times\Sigma\to M$, 
$\fnb_\varsigmaD\alpha\in\GammaSD\Lambda\Sigma$,
$\fnb\colon N\hookrightarrow\Sigma$,
$\alpha\in\Gamma\Lambda^{\dim N}N$ and   
$\Psi \in\GammaSD\Lambda^{\dim M-1}M$ be as in theorem
\ref{thm_Transp_initial_distr}. Then
\begin{equation}
\Psi[\phi]=\int_{\tau\in\Real} 
\Phi[\zeta_\tau^\star i_v\phi]\, d\tau
\label{Trans_int_soln}
\end{equation}
for all $\phi\in\Gamma_0\Lambda^1 M$ where
$\Phi=\fnb_\varsigmaD\alpha$ and $\zeta_\tau:\Sigma\to M$,
$\zeta_\tau(y)=\fnc(\tau,y)$. 
\end{corrol}
\begin{proof}
The pushforward of the vector 
$\partial_\tau|_{(\tau,y)}\in T_{(\tau,y)}(\Real\times N)$ under
$\Real\times N\stackrel{\sId\times\sfnb}{\longrightarrow}
\Real\times\Sigma\stackrel{\sfnc}{\to} M$ is given by
$(\Id\times\fnb)_\star(\partial_\tau|_{(\tau,y)})=
\partial_\tau|_{(\tau,\sfnb(y))}$ and
$\fnc_\star\partial_\tau|_{(\tau,\sfnb(y))} = v|_{\fnc(\tau,\sfnb(y))}$.

Let $I_\tau:N\to\Real\times N$, $I_\tau(y)=(\tau,y)$ so that 
$(c\circ (\Id\times\fnb)\circ I_\tau)(y)=
c((\Id\times\fnb)(\tau,y))=c(\tau,\fnb(y))=\zeta_\tau(\fnb(y))$. Thus
\begin{eqnarray*}
\fl
\Psi[\phi]
&=
\fna_\varsigmaD(\pr_2^\star\alpha)[\phi]
=
\DD(\pr_2^\star\alpha)[\fna^\star\phi]
=
\int_{\Real\times N} \fna^\star\phi\wedge\pr_2^\star\alpha
=
\int_{\Real\times N} (\Id\times\fnb)^\star(\fnc^\star \phi)
\wedge\pr_2^\star\alpha
\\
\fl&=
\int_{\tau\in\Real}\!d\tau\! \int_{N} I_\tau^\star i_{\partial\tau}
\Big((\Id\times\fnb)^\star(\fnc^\star \phi)
\wedge\pr_2^\star\alpha\Big)
=
\int_{\tau\in\Real}\!d\tau\!\! \int_{N} I_\tau^\star 
\Big((\Id\times\fnb)^\star i_{\partial\tau}(\fnc^\star \phi)
\wedge\pr_2^\star\alpha\Big)
\\\fl&=
\int_{\tau\in\Real}\!d\tau\! \int_{N} I_\tau^\star 
\Big((\Id\times\fnb)^\star (\fnc^\star i_v\phi)
\wedge\pr_2^\star\alpha\Big)
=
\int_{\tau\in\Real}\!d\tau\! \int_{N} 
(c\circ (\Id\times\fnb)\circ I_\tau)^\star (i_v\phi)
\wedge\alpha
\\\fl&=
\int_{\tau\in\Real}\!d\tau\! \int_{N} 
\fnb^\star\big(\zeta_\tau^\star (i_v\phi)\big)
\wedge\alpha
=
\int_{\tau\in\Real}d\tau 
\fnb_\varsigmaD(\alpha)[\zeta_\tau^\star (i_v\phi)\big]
=
\int_{\tau\in\Real}d\tau 
\Phi[\zeta_\tau^\star (i_v\phi)\big]
\end{eqnarray*}

\end{proof}


\subsection{Distributional solutions to the Liouville equations as a source
  for Maxwell's equations.}
\label{ch_Trans_source}

The following shows that the source for Maxwell's equations is closed.
\begin{theorem}
\label{thm_Transp_source_closed}
The source for Maxwell's equations (\ref{MV_Maxwell}) is closed
i.e. $d\Jsource=0$.
\end{theorem}
\begin{proof}
If $\pdd$ is bounded with respect to $\pi$ then lemma
\ref{lm_Push_d_comm_bdd} implies
$d(\pi_\varsigma\pdd)=\pi_\varsigma(d\pdd)$.  If $\pdd$ is regular and
unbounded but $\pi_\varsigma\pdd$ is defined then in \cite{BottTu} it
is also shown that $d\circ\pi_\varsigma=\pi_\varsigma\circ d$.  Thus
\begin{equation*}
d\Jsource
=
d (\pi_\varsigma \pdd)
=
\pi_\varsigma (d \pdd)
=
0
\end{equation*}
\end{proof}

We have established that a submanifold distributional solution to the
transport equations can be constructed from a distribution on an
initial hypersurface.  From now on we consider distributional
solutions for the transport equations (\ref{Trans_transp_eqns}) for
distributions $\Psi\in\GammaSD\Lambda E$ where $\pi\colon {E}\to M$ is
a fibre bundle.  If this solution is to be the source for an
electromagnetic field then we need to be able to take its pushforward
with respect to the projection map $\pi$. However this projection map
is not proper. Therefore we must establish under what condition we can
guarantee that $\Psi$ is bounded with respect to $\pi$.
\begin{lemma}
\label{lm_Trans_source_bdd}
Let $\fna\colon N\hookrightarrow{E}$ and
$\fna_\varsigmaD(\alpha)\in\GammaSD\Lambda{E}$ be bounded with
respect to $\pi\colon {E}\to M$ then $\DD(\alpha)$ is bounded with respect
to $(\pi\circ \fna)\colon N\to M$ and 
\begin{equation}
\pi_\varsigma(\fna_\varsigmaD\alpha)=(\pi\circ \fna)_\varsigmaD\alpha
\label{Trans_source_pi_a}
\end{equation}
\end{lemma}
\begin{proof}
Let $U\subset M$ be compact. Since $\fna_\varsigmaD(\alpha)$ is
bounded with respect to $\pi$, it follows that
$\pi^{-1}U\inter\supp(\fna_\varsigmaD\alpha)$ is compact. As $\fna$ is
proper $\fna^{-1}\big(\pi^{-1}U\inter\fna(N))\big)\subset N$ is
compact. Now
\begin{equation*}
\fl
\fna^{-1}\big(\pi^{-1}U\inter \fna(N)\big)
=
(\fna^{-1}\pi^{-1}U)\inter (\fna^{-1}\fna(N))
=
(\pi\circ \fna)^{-1} U\inter N
=
(\pi\circ \fna)^{-1} U\inter \supp(D\alpha)
\end{equation*}
Thus $\DD(\alpha)$ is bounded with respect to $(\pi\circ \fna)$.
Given $\phi\in\Gamma_0\Lambda M$ and $h\in\Gamma_0\Lambda E$ satisfies
(\ref{Push_bdd_h_req}) i.e
$\supp(1-h)\inter\supp(\fna^\star\phi)\inter\supp(\fna_\varsigmaD\alpha)=
\emptyset$ then we wish to show that $\fna^\star h\in\Gamma_0\Lambda
N$ satisfies (\ref{Push_bdd_h_req}) for $\Psi=\DD\alpha$ and $\phi$
replaced by $\pi^\star\phi$:
\begin{eqnarray*}
\fl\qquad
\lefteqn{\supp(1-\fna^\star h) \inter 
\supp(\fna^\star\pi^\star\phi)\inter
\supp(\DD\alpha)}\qquad&
\\\fl
&=
\supp(1-\fna^\star h) \inter 
\supp(\fna^\star\pi^\star\phi)\inter N
=
\supp(\fna^\star (1-h)) \inter 
\supp(\fna^\star\pi^\star\phi)
\\\fl&
\subset
\fna^{-1}\big(\supp(1-h)\big) \inter 
\fna^{-1}\big(\supp(\pi^\star\phi)\big)
=
\fna^{-1}\big(\supp(1-h) \inter 
\supp(\pi^\star\phi)\big)
\\\fl&
=
\fna^{-1}\big(\supp(1-h) \inter 
\supp(\pi^\star\phi)\inter \fna(N)\big)
=
\fna^{-1}\big(
\supp(1-h)\inter\supp(\fna^\star\phi)\inter\supp(\fna_\varsigmaD\alpha)
\big)
=\emptyset
\end{eqnarray*}
Hence we have
\begin{eqnarray*}
\fl\qquad
\pi_\varsigma(\fna_\varsigmaD\alpha)[\phi]
&=
\fna_\varsigmaD\alpha[h\,\pi^\star\phi]
=
\DD\alpha[\fna^\star(h\,\pi^\star\phi)]
=
\DD\alpha[\fna^\star(h)\,(\fna^\star\pi^\star\phi)]
\\\fl&=
\DD\alpha[\fna^\star(h)\,(\pi\circ\fna)^\star(\phi)]
=
(\pi\circ\fna)_\varsigmaD\alpha[\phi]
\end{eqnarray*}
giving (\ref{Trans_source_pi_a}).
\end{proof}

Recall that a vector field $\LiouV\in\Gamma T E$ with $\pi\colon E\to
M$ and $E\subset TM$ is \defn{horizontal} if $\pi_\star(\LiouV|_u)=u$
for all $u\in E$. The following lemma relates the integral curves of 
horizontal vector fields with curves on the base space $M$.
\begin{lemma}
\label{lm_horizontal}
Let $\pi\colon E\to M$ with $ E\subset T M$ and $\LiouV\in\Gamma T E$
be a horizontal vector field and let $\gamma\colon\Real\to E$ be an integral
curve of $\LiouV$ then $\gamma=\Cd$ where
$C=\pi\circ\gamma\colon\Real\to M$.
\end{lemma}
\begin{proof}
Let $\tau\in\Real$. Recall that $\partial_\tau\in T_\tau\Real$ is the
natural vector field and $\partial_\tau|_\tau\in\Gamma T_\tau\Real$ is a
vector at the point $\tau\in\Real$. The lifts of the curves
$\dot\gamma=\gamma_\star(\partial_\tau)\colon\Real\to T E$ and 
$\Cd=C_\star(\partial_\tau)\colon\Real\to T M$. 
\begin{equation*}
\fl\qquad
\Cd(\tau)
=
C_\star(\partial_\tau|_\tau)
=
(\pi\circ\gamma)_\star(\partial_\tau|_\tau)
=
\pi_\star\big(\gamma_\star(\partial_\tau|_\tau)\big)
=
\pi_\star\big(\dot\gamma(\tau)\big)
=
\pi_\star\big(\LiouV_{\gamma(\tau)}\big)
=
\gamma(\tau)
\end{equation*}

\end{proof}

We can ask the question: 
\begin{equation}
\parbox{0.7\textwidth}{
If given an initial hypersurface
$\sigma\colon \Sigma\hookrightarrow{E}$, with respect to a horizontal
vector field, such that $\sigma^\varsigma\Psi$
is bounded with respect to $\pi\circ\sigma$ then is $\Psi$ bounded
with respect to $\pi$?
}
\label{Trans_source_question}
\end{equation}
\vspace{0.5em}

\begin{figure}
\centerline{
\setlength{\unitlength}{0.04\textwidth}
\begin{picture}(10,6.5)
\put(0,0){\includegraphics[width=0.4\textwidth,viewport=237 383 546 575
]{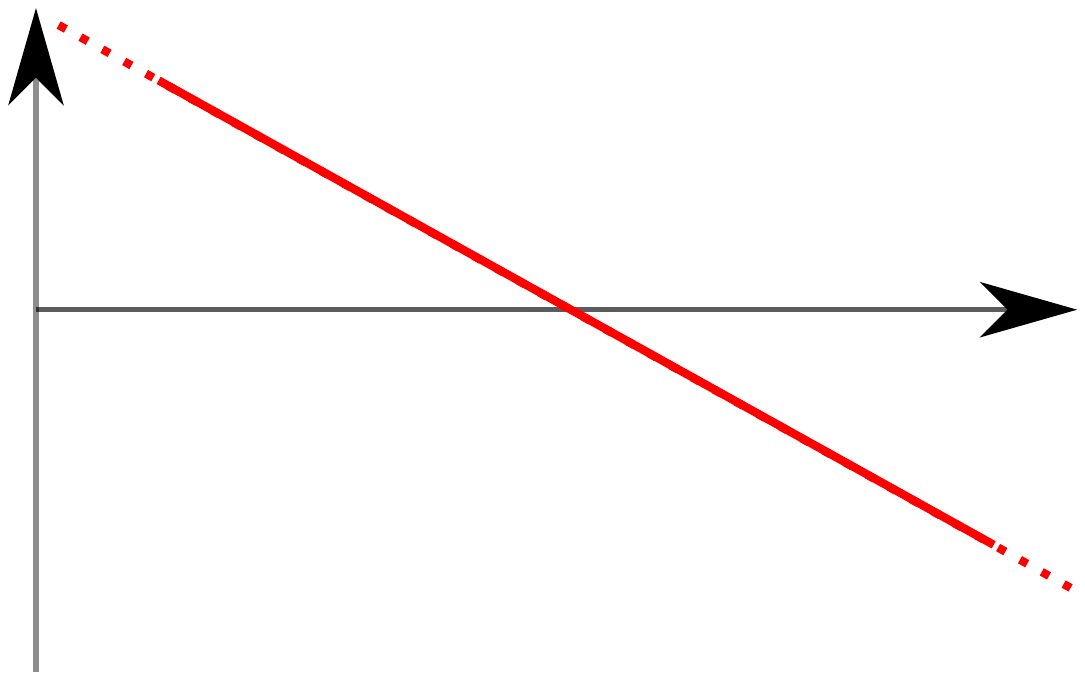}}
\put(-0.6,5){$\xdot^1$}
\put(8,2.7){$x^1$}
\put(8,5){$x^0<0$}
\end{picture}
\qquad\qquad
\begin{picture}(10,6)
\put(0,0){\includegraphics[width=0.4\textwidth,viewport=237 383 546 575
]{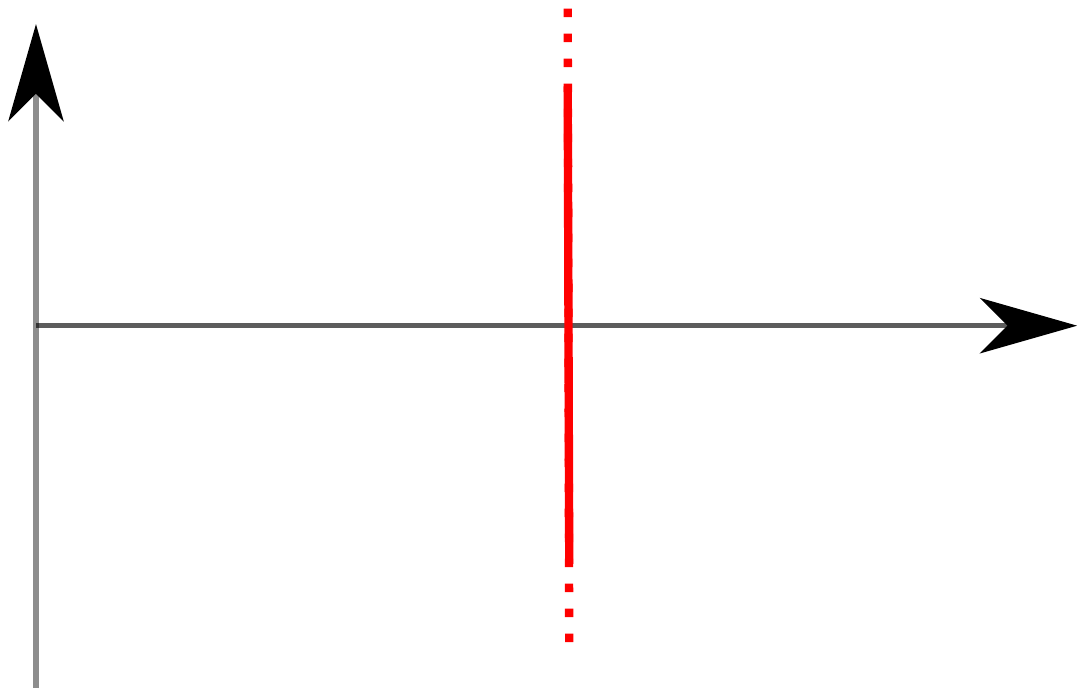}}
\put(-0.6,5){$\xdot^1$}
\put(8,2.7){$x^1$}
\put(8,5){$x^0=0$}
\end{picture}
}
\caption{Distribution of charge for a Galilean spacetime for two
  different values of $x^0$.}
\label{fig_Example1}
\end{figure}

It turns out that even for simple cases this is not the case.  For
example consider the Vlasov operator corresponding to a simple force
free drift in one dimensional Galilean spacetime $M=\Real^2$
coordinates $(t,x)$ with ${E}=\Real^3$ coordinated by $(t,x,\xdot)$
and $\pi(t,x,\xdot)=(t,x)$.
\begin{equation}
\LiouV=\pfrac{}{t}+\xdot\pfrac{}{x}
\end{equation}
A solution to the transport equations for this vector field is given
by $\Psi=\fna_\varsigmaD \alpha\in\GammaSD\Lambda^2{E}$ where
$\fna\colon \Real^2\to{E}$, $\fna(\tau,y)=(\tau,\tau y,y)$ and
$\alpha=\hat\alpha(y)d y\in\Gamma\Lambda^1\Real^2$.

Now to see that $\Psi$ is a solution to the transport equations we use
theorem \ref{lm_Trans_soln}. Clearly $d\alpha=0$ and
\begin{equation*}
\fna_\star\Big(\pfrac{}{\tau}\Big)\Big|_{(\tau,y)}=\pfrac{}{t}+y\pfrac{}{x}
=\LiouV|_{\fna(\tau,y)}
\end{equation*}
Hence it is a solution. 

With initial hypersurface $\sigma\colon \Sigma\to{E}$,
$\sigma(x_0,\xdot_0)=(t_0,x_0,\xdot_0)$ an initial distribution is
$\sigma^\varsigma\Psi=\fnb_\varsigmaD\hat\alpha$ where 
$\fnb\colon \Real\to\Sigma$, $\fnb(y)=(t_0y,y)$, so that
$\sigma(\fnb(y))=(t_0,t_0y,y)$. Given compact $U\subset M$ we have
\begin{eqnarray*}
\fl\qquad
\supp(\sigma^\varsigma\Psi)\inter \pi^{-1}U
&=
\supp(\fnb_\varsigmaD\hat\alpha)\inter \pi^{-1}U
=
\Set{(t_0,t_0y,y)\big|y\in\Real}\inter
\Set{(t,x,\xdot)\big|(t,x)\in U}
\\&=
\Set{(t_0,t_0y,y)\big|(t_0,t_0y)\in U} 
\end{eqnarray*}
is compact as long as $t_0\ne0$. Thus $\sigma^\varsigma\Psi$
is bounded with respect to $\pi\circ\sigma$, see figure
\ref{fig_Example1}. 
However consider $(0,0)\in M$ then
\begin{equation*}
\fl\qquad
\pi^{-1}\Set{(0,0)}=\Set{(0,0,\xdot)|\xdot\in\Real}
=\Set{(0,0,y)|y\in\Real}=\Set{\fna(0,y)|y\in\Real}
\subset\supp(\Psi)
\end{equation*}
Thus $\Psi$ is not bounded with respect to $\pi$.

\vspace{1em}

It turns out that for the relativistic Maxwell-Vlasov equations, as
long as the initial hypersurface $\Sigma\subset{E}$ is such that
$\pi(\Sigma)\subset M$ is a Cauchy surface, then we can guarantee that
if $\sigma^\varsigma\Psi$ is bounded with respect to $\pi\circ\sigma$
then $\Psi$ is bounded with respect to $\pi$.

\begin{theorem}
\label{thm_Transp_source_init}
Let $\pi\colon {\Ebun}\to \Man$ be the upper unit hyperboloid bundle over a
globally hyperbolic spacetime. Let $\LiouV\in\Gamma T{\Ebun}$ be a horizontal
vector field, i.e. such that $\pi_\star(\LiouV|_u)=u$. Let
$\sigma\colon \Sigma\hookrightarrow{\Ebun}$ be an initial hypersurface such
that $\pi(\sigma(\Sigma))\subset \Man$ is a Cauchy surface.  Let
$\Psi\in\GammaSD\Lambda{\Ebun}$ be a solution to the transport equations
with respect to $\LiouV$. Let $\sigma^\varsigma(\Psi)$ be bounded with
respect to $\pi\circ\sigma$ then $\Psi$ is bounded with respect to
$\pi$.
\end{theorem}
\begin{proof}
Let $\SigmaHat=\pi(\sigma\Sigma)$,
$\hat\sigma\colon\SigmaHat\hookrightarrow\Man$ and
$\hat\pi=\pi\circ\sigma:\Sigma\to\SigmaHat$. Given a compact subset
$Y\subset \Man$, since $\SigmaHat$ is a Cauchy surface, it is shown in
\cite{ONeil}
that for each $y\in Y$ the light cone of $y$ intersecting with
$\SigmaHat$ is compact. Define $X\subset\SigmaHat$ as
\begin{equation*}
\fl\quad
X=\Set{x\in \SigmaHat\,\Big|\,
\textup{there exists a timelike or lightlike curve $C$ passing though $x$
  and $Y$}} 
\end{equation*}
then $X$ is compact.
Furthermore since $X$ and $Y$ are compact the proper time it takes to go from
$X$ to $Y$ has a maximum and minimum.
\begin{equation*}
T=\Set{
\tau\in\Real\,\bigg|\,
\textup{\begin{tabular}{c}
there exists a timelike normalised curve
  $C\colon\Real\hookrightarrow\Man$\\ such that $C(0)\in \hat\sigma(X)$
  and $C(\tau)\in Y$
      \end{tabular}
}
}
\end{equation*}
Then $T\subset\Real$ is compact. We have the following commutative
diagram
\begin{equation*}
\xymatrix{ 
N \arhook[r]^{I_0} \arhook[d]_{\sfnb} &
\Real\times N \arhook[d]^{\sfna} 
\\
\Sigma \arhook[r]^{\sigma} \ar[d]_{\hat\pi} &
\Ebun \ar[d]_{\pi}
\\
\SigmaHat \arhook[r]^{\hat\sigma} &
\Man
}
\end{equation*}
Since $\sigma^\varsigma\Psi=\fnb_\varsigmaD\alpha$ is bounded with
respect to $\hat\pi$, and $X\subset\SigmaHat$ is compact then
$\hat\pi^{-1}(X)\inter\supp(\fnb_\varsigmaD\alpha)\subset\Sigma$ is
compact. Now $\hat\pi^{-1}(X)\inter\supp(\fnb_\varsigmaD\alpha)=
\hat\pi^{-1}(X)\inter\fnb(N)$ and since $\fnb\colon N\to\Sigma$
is proper 
\begin{equation*}
\fnb^{-1}\big(\hat\pi^{-1}(X)\inter\fnb(N)\big)=
\fnb^{-1}\hat\pi^{-1}(X)\inter N=
\fnb^{-1}\hat\pi^{-1}(X)\subset N
\quad\textup{is compact}
\end{equation*}

We need to show that $\pi^{-1}(Y)\inter\supp(\Psi)$ is compact. 
Now from set theory and the fact that $\fna$ is injective and
$\supp(\Psi)=\fna(N)$ then
$\pi^{-1}(Y)\inter\supp(\Psi)=\fna(\fna^{-1}\pi^{-1}(Y))$. Now since
$\fna$ is continuous, it is sufficient to show that
$\fna^{-1}\pi^{-1}(Y)\subset\Real\times N$ is compact.

Let $(\tau_0,y)\in \fna^{-1}\pi^{-1}(Y)$, so that
$\pi\big(\fna(\tau_0,y)\big)\in Y$. Let $\gamma\colon\Real\to\Ebun$ be given by
$\gamma(\tau)=\fna(\tau,y)$, then from 
(\ref{Trans_soln_def_S}) $\gamma$ is an integral curve of
$\LiouV$. From lemma \ref{lm_horizontal}, $\gamma=\Cd$ where
$\Cd=\pi\circ\gamma$. Since $\Cd\colon\Real\to\Ebun$,
$g(\Cd,\Cd)=-1$ so $C$ is a timelike normalised curve.
Now
\begin{equation*}
C(0)=\pi\big(\gamma(0)\big)=
\pi\big(\fna(0,y)\big)=
\pi\big(I_0(y)\big)=
\pi\big(\sigma(\fnb(y))\big)=
\hat\sigma\big(\hat\pi(\fnb(y))\big)
\end{equation*}
thus $\hat\pi(\fnb(y))\in\hat\sigma(\SigmaHat)$. Also
$C(\tau_0)=\pi(\gamma(\tau_0))=\pi(\fna(\tau_0,y))\in Y$.  From the
definitions of $X$ and $T$, $C(0)\in \hat\sigma(X)$ and $\tau_0\in
T$. Since $C(0)=\hat\sigma\big(\hat\pi\big(\fnb(y)\big)\big)\in
\sigma(X)$ then $\hat\pi\big(\fnb(y)\big)\in X$ so 
$y\in\fnb^{-1}\hat\pi^{-1}(X)$. Hence $(\tau_0,y)\in T\times
\fnb^{-1}\hat\pi^{-1}(X)$. Thus we have shown that
\begin{equation*}
\fna^{-1}\pi^{-1}(Y)\subset T\times\fnb^{-1}\hat\pi^{-1}(X)
\end{equation*}
Being the product of two compact sects $T\times\fnb^{-1}\hat\pi^{-1}(X)$ is
compact and since $\fna^{-1}\pi^{-1}(Y)$ is closed it is therefore
compact. Hence result.
\end{proof} 

The condition that $\pi(\sigma(\Sigma))=\SigmaHat\subset \Man$ is a Cauchy
surface is required. Here we give an example where $\Man$ is globally
hyperbolic, $\sigma^\varsigma(\Psi)$ is bounded with respect to
$\pi\colon \Ebun\to\Man$ but $\Psi$ is not bounded with respect to $\pi$.

\begin{figure}
\centerline{
\setlength{\unitlength}{0.04\textwidth}
\begin{picture}(10,8.5)
\put(0,0){\includegraphics[width=0.328\textwidth,viewport=112 235 411
  538]{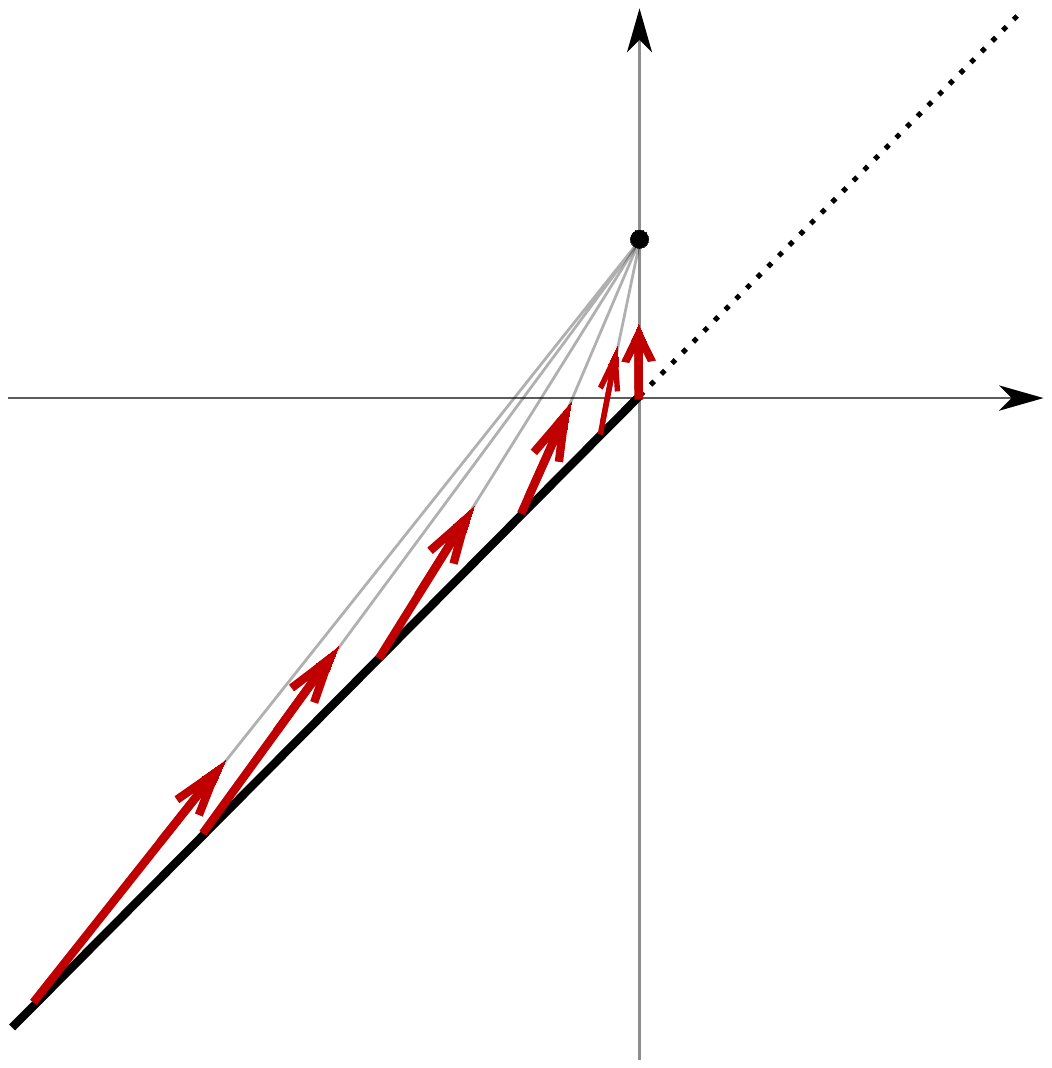}}
\put(4.3,7.5){$x^0$}
\put(7.5,4.5){$x^1$}
\put(1,0.8){\rotatebox{45}{\small Initial Distribution}}
\put(0.4,1.2){\rotatebox{57}{\small Initial Vectors}}
\put(6.3,6){\rotatebox{45}{$\SigmaHat$}}
\put(3.8,6.5){\footnotesize $(0,1)$}
\end{picture}
}
\caption{On the lightlike initial hypersurface, all initial velocities
  point towards the point $(0,1)$}
\label{fig_Example2}
\end{figure}

Consider $\Man$ is two dimensional Minkowski spacetime, with
coordinates $(x^0,x^1)$ and $\Ebun$ is the upper unit hyperboloid
coordinates $(x^0,x^1,y)$. Let $\LiouV$ be the Liouville vector
(\ref{MV_def_W}) for a zero electric force:
\begin{equation*}
\LiouV = 
\sqrt{1+y^2} \pfrac{}{x^0} + y\pfrac{}{x^1} 
\end{equation*}
Let 
\begin{equation*}
\sigma\colon \Sigma\hookrightarrow\Ebun\;,\qquad
\sigma(z,p)=(x^0=z,x^1=z,y=p)
\end{equation*}
so that $\pi\circ\sigma\colon \Sigma\to\Man$, 
$\pi\circ\sigma(z,p)=(z,z)$ is a lightlike curve so $\sigma(\Sigma)$
must be transverse to $\LiouV$. Also every integral curve of $\LiouV$
must intersect $\sigma(\Sigma)$ so  $\Sigma$ is an initial hypersurface
of $\LiouV$. Clearly $\pi(\sigma\Sigma)$ is not a Cauchy
surface. Let $N=\Real\times\Set{\zhat\in\Real,\zhat\le 0}$ and 
\begin{equation*}
\fl\qquad
\fna\colon N\hookrightarrow\Ebun\,,\qquad
\fna(\tau,\zhat)=\Big(
x^0=\frac{1-\zhat}{\sqrt{1-2\zhat}}\tau + \zhat,
x^1=\frac{-\zhat}{\sqrt{1-2\zhat}}\tau + \zhat,
y=\frac{-\zhat}{\sqrt{1-2\zhat}}
\Big)
\end{equation*}
so that 
\begin{equation*}
\fna_\star\Big(\pfrac{}{\tau}\Big)=\Big(
\frac{1-\zhat}{\sqrt{1-2\zhat}},
\frac{-\zhat}{\sqrt{1-2\zhat}}, 0 \Big)
=
\LiouV|_{\fna(\tau,\zhat)}
\end{equation*}
Hence $\fna$ is tangential to $\LiouV$. Let $\Psi=\fna_\varsigmaD1$. 
Now $\sigma^\varsigma(\Psi)=\fnb_\varsigmaD 1$ where
\begin{equation*}
\fnb\colon \Set{\zhat\in\Real,\zhat\le 0}\to\Sigma\,,\qquad
\fnb(\zhat)=\big(\zhat,\zhat,{-\zhat}({1-2\zhat})^{-1/2}\big)
\end{equation*}
This map is proper and $\sigma^\varsigma(\Psi)$ is bounded with respect
to $\pi$. However this example has been contrived so that every
integral curve passes through $(x^0,x^1)=(1,0)\in\Man$. See figure
\ref{fig_Example2}.
\begin{equation*}
\fna(\sqrt{1-2\zhat},\zhat)=\big(1,0,{-\zhat}({1-2\zhat})^{-1/2}\big) 
\end{equation*}
So 
\begin{equation*}
\pi^{-1}\Set{(1,0)}\inter\supp(\Psi)=
\Set{(1,0,y)|y\ge0}
\end{equation*}
is not compact so $\Psi$ is not bounded with respect to $\pi$.


\section{Known solutions to the Maxwell-Vlasov equations.}
\label{ch_Examples}

\subsection{Worldline of a point charge and the Klimontovich distribution}
\label{ch_point}

For the worldline of a point charge we can show that the Liouville
equation (\ref{MV_pdd_closed},\ref{MV_Liou_eqn}) implies that the
point charge undergoes the Lorentz force equation
(\ref{Lift_force_eqn}) and that the electromagnetic source due to a
point charge is given by $\Jsource=C^D\in\GammaSD\Lambda^3\Man$ where
$C^D=C_\varsigmaD(1)$. However due to the divergence of the
electromagnetic field $F$ due to point sources we cannot demand that
there exist solutions to the complete Maxwell-Vlasov equation
(\ref{MV_def_Jsource}-\ref{MV_Liou_eqn}). It is therefore necessary to
perform a regularisation of $F$ before substituting into
(\ref{MV_def_W}). For a single point charge the usual regularisation
is the Dirac regularisation which lads to the Lorentz-Dirac equation.

Let $C\colon \Real\to \Man$ be a regular curve with $\metric(\Cd,\Cd)=-1$ 
and let $\Cd\colon \Real\to\Ebun$ be the corresponding lift. Let
\begin{equation}
\fl\qquad
\pdd\in\GammaD\Lambda^6\Ebun
\;;\quad
\pdd=\Cd_\varsigmaD 1 = \Cd^D
\quadtext{i.e.}
\pdd[\phi]= \int_{\Real} \Cd^\star (\phi) 
\quadtext{for all}
\phi\in\Gamma_0\Lambda^1\Ebun
\label{point_rho}
\end{equation}
Clearly $d\pdd=0$.

\begin{lemma}
Equations (\ref{MV_def_Jsource}) and
(\ref{point_rho}) give the point
source.
\begin{equation}
\Jsource = C^D = C_\varsigmaD 1
\label{point_Maxwell}
\end{equation}

\end{lemma}
\begin{proof}
$\Cd$ is bounded with respect to $\pi$ thus from lemma
\ref{lm_Trans_source_bdd} we have
\begin{equation*}
\Jsource 
=
\pi_\varsigma (\pdd)
=
\pi_\varsigma(\Cd_\varsigmaD 1)
=
(\pi\circ\Cd)_\varsigmaD 1
=
C_\varsigmaD 1
=
C^D
\end{equation*}
\end{proof}

\begin{lemma}
Equations (\ref{MV_def_W}), (\ref{MV_Liou_eqn}) and
(\ref{point_rho}) imply that the point charge
undergoes the Lorentz force equation:
\begin{equation}
\nabla_\Cd \Cd = \dual{i_\Cd F}
\label{point_Lorentz_force}
\end{equation}
\end{lemma}
\begin{proof}
From (\ref{MV_Liou_eqn}) and (\ref{point_rho}) we have
$i_\LiouV\Cd^D=0$ hence from lemma \ref{lm_intg_surf} we have $\Cd$ is
tangential to $\LiouV$ hence $\Cd$ is an integral curve of $\LiouV$
with a scaling. I.e.  $\LiouV|_{\Cd(\tau)}=\kappa(\tau)\Fmul
\Cdd(\tau)$.  By hitting both sides with $\pi_\star$ we see
\begin{equation*}
\Cd(\tau)=
\pi_\star \LiouV|_{\Cd(\tau)}
=
\pi_\star \big(\kappa\Fmul \Cdd(\tau)\big)
=
\kappa\Fmul\Cd(\tau)
\end{equation*}
hence $\kappa=1$. Thus $\LiouV|_{\Cd(\tau)}=\Cdd(\tau)$. So $\Cd$ is
an integral curve of $\LiouV$ and thus
(\ref{Lift_force_eqn}) gives (\ref{point_Lorentz_force}).
\end{proof}

As stated in the introduction the Klimontovich distribution is
simply a finite number of point worldlines. Thus we can set
\begin{equation*}
\Theta=\sum_{k=1}^N \Cd^D_k
\end{equation*}
Which gives the source for Maxwell's equation as
\begin{equation*}
\Jsource=\sum_{k=1}^N C^D_k
\end{equation*}
Again, this does not give rise to a continuous electromagnetic field
so some method of regularisation is necessary. For the Klimontovich
distribution, for example, we can ignore the contribution of $C_j$ to
$\Jsource$ when calculating the motion of the worldline $\Cd_j$. That
is we solve $d (\DD F_j)=0$ and $d (\DD(\star  F_j))=\sum_{k\ne j} C^D_k$. We
then construct the Liouville vector field $\LiouV_j=\LiouV(F_j)$ using
(\ref{MV_def_W}) and solve the transport equations for $\Cd_j$ using
$\LiouV_j$.

\subsection{Cold charged fluid model}
\label{ch_cold}

Here we see that the cold charged fluid model\cite{bgt} is simply an
example of a distributional solution to the Maxwell-Vlasov equations.
Let $v\in\Gamma T\Man$ with $\metric(v,v)=-1$ and
$\rho\in\Gamma\Lambda^0 \Man$.  We may consider $v\colon
\Man\hookrightarrow\Ebun$ as a closed embedding.  Let
\begin{eqnarray}
&\pdd\in\GammaD\Lambda^6\Ebun
\;;\qquad
\pdd=v_\varsigmaD(\rho\star\dual{v})
\label{cold_rho}
\\
&\qquadtext{i.e.}
\pdd[\phi]= \int_{x\in \Man} \phi|_{(v|_x)} \wedge\rho|_x\star \dual{v}|_x
\quadtext{for all}
\phi\in\Gamma_0\Lambda^1\Ebun
\nonumber
\end{eqnarray}
where $\dual{v}\in\Gamma\Lambda^1\Man$ is the metric dual of $v$
satisfying $\dual{v}(u)=g(v,u)$ for all $u\in\Gamma T\Man$.

\begin{lemma}
Equations (\ref{MV_def_Jsource}) and (\ref{cold_rho})
give the source for the cold charged fluid:
\begin{equation}
\Jsource = \DD(\rho\star \dual{v})
\label{cold_Maxwell}
\end{equation}
\end{lemma}
\begin{proof}
$\pdd$ is bounded with respect to $\pi$ thus from lemma
\ref{lm_Trans_source_bdd} we have
\begin{equation*}
\Jsource
=
\pi_\varsigma \pdd
=
\pi_\varsigma(v_\varsigmaD(\rho\star\dual{v}))
=
(\pi\circ v)_\varsigmaD(\rho\star\dual{v})
=
\DD(\rho\star\dual{v})
\end{equation*}

\end{proof}
\noindent
Also since $d\Jsource=0$ we have the continuity equation 
\begin{equation}
d (\rho\star\dual{v}) = 0
\label{cold_continuety}
\end{equation}

\begin{lemma}
Equations 
(\ref{MV_def_W}),(\ref{MV_Liou_eqn}) and (\ref{cold_rho})
imply the Lorentz force equation
\begin{equation}
\nabla_v v = \dual{i_v F}
\label{cold_Lorentz_force}
\end{equation}
\end{lemma}
\begin{proof}
From (\ref{MV_Liou_eqn}) and (\ref{cold_rho}) we have $i_\LiouV
\big(v_\varsigmaD(\rho\star\dual{v})\big)=0$. Thus from lemma
\ref{lm_intg_surf} there exists $u\in\Gamma T\Man$ such that
$v_\star(u|_x)=\LiouV|_{v|_x}$. Thus $u|_x=\pi_\star
v_\star(u|_x)=\pi_\star \LiouV|_{v|_x}=v|_x$ hence $u=v$.  Hence
$v_\star(v|_x)=\LiouV|_{v|_x}$. 

Given an integral curve $C\colon \Real\to\Man$ of $v$ then
$v|_{C(\tau)}=\Cd(\tau)$, i.e. $v\circ C=\Cd$. Thus
\begin{eqnarray*}
\Cdd(\tau_0) &= 
\Cd_\star (\partial_\tau|_{\tau_0}) = 
(v\circ C)_\star (\partial_\tau|_{\tau_0}) = 
v_\star \big(C_\star (\partial_\tau|_{\tau_0})\big) = 
v_\star \big(\Cd(\tau_0)\big) \\&= 
v_\star \big(v|_{C(\tau_0)}\big) = 
\LiouV|_{v|_{C(\tau_0)}}=
\LiouV|_{\Cd(\tau_0)}
\end{eqnarray*}
Hence $\Cd$ is an integral curve of $\LiouV$ and hence from 
(\ref{Lift_force_eqn}) $\nabla_\Cd \Cd = \dual{i_\Cd F}$. Since this
is true for all integral curves we have (\ref{cold_Lorentz_force}).

\end{proof}

\subsection{Multicurrent model}
\label{ch_fold}

Here we see that the multicurrent model is simply an example of a
distributional solution to the Maxwell-Vlasov equations. The
conclusions
(\ref{fold_Maxwell_pullback},\ref{fold_Maxwell_intg},\ref{fold_Lorentz_force})
are in \cite{GIFT2,folds2}, with slightly modified notation. An
application to wake breaking in plasmas is given in \cite{LiWo}.

Let $\Bman=\Real\times\underline \Bman$ be a body-time manifold with
points $(\tau,\Vy)\in \Bman$ and a measure
$d\tau\wedge\Jform\in\Gamma\Lambda^4\Bman$ where
$\Jform=\pi_2^\star(\underline\Jform)\in\Gamma\Lambda^3\Bman$ and
$\underline\Jform\in\Gamma\Lambda^3\underline \Bman$. Let $C\colon
\Bman\to \Man$ with $\Cd=C_\star(\partial_\tau)\in\Ebun$. Let
\begin{eqnarray}
&\pdd\in\GammaD\Lambda^6\Ebun
\;;\qquad
\pdd=\Cd_\varsigmaD(\Jform)
\label{fold_rho}
\\
&\qquadtext{i.e.}
\pdd[\phi]= \int_{(\tau,\Vy)\in \Bman}
\phi|_{\Cd(\tau,\Vy)} 
\wedge\Jform
\quadtext{for all}
\phi\in\Gamma_0\Lambda^1\Ebun
\nonumber
\end{eqnarray}

\begin{lemma}
\label{lm_fold_push}
Equations (\ref{MV_def_Jsource}) and (\ref{fold_rho}) give the source
$\Jsource$ due to a multicurrent fluid. These may be written in a variety
of ways. Using distributional push forwards:
\begin{equation}
\Jsource = C_\varsigmaD(\Jform)
\label{fold_Maxwell_push}
\end{equation}
Furthermore if $\Jsource$ is regular in the sense that it is piecewise
continuous, then we can write $\Jsource=\DD(\hat\Jsource)$ where
$\hat\Jsource\in\Gamma_{\textups{pc}}\Lambda^3\Man$ is piecewise
continuous. 

In terms of the inverse pull back on generic open sets
\begin{equation}
\hat\Jsource
= 
\sum_{i=1}^{N(U^\Man)}\textup{sign}(\det(C_{[i]\star}))\Fmul 
C_{[i]}^{-1\star}(\Jform)
\label{fold_Maxwell_pullback}
\end{equation}
where $U^\Man\subset \Man$ is an open set where $C$ is generic and therefore
the number of preimages $U^\Bman_{[i]}$, $i=1,\ldots,N(U^\Man)$ is constant,
and the maps $C_{[i]}\colon U^\Bman_{[i]}\to U^\Man$ are invertible.
In terms of integrals over spacelike hypersurfaces of $\Man$.
\begin{equation}
\fl\qquad\qquad
\int_S \hat\Jsource = \int_{C^{-1}(S)} \Jform
\quadtext{for all bounded spacelike hypersurfaces}
S\subset \Man
\label{fold_Maxwell_intg}
\end{equation}
\end{lemma}
\begin{proof}
Since $\Cd(\tau,\Vy)=C_\star(\partial_\tau|_{(\tau,\Vy)})$,
\begin{equation*}
\pi(\Cd(\tau,\Vy))=
\pi
C_\star(\partial_\tau|_{(\tau,\Vy)})
=
C(\tau,\Vy)
\end{equation*}
i.e. $\pi\circ\Cd=C$.
Hence from lemma \ref{lm_Trans_source_bdd}
\begin{equation*}
\Jsource=\pi_\varsigma(\Cd_\varsigmaD\Jform)=
(\pi\circ\Cd)_\varsigmaD(\Jform)=
C_\varsigmaD(\Jform)
\end{equation*}
giving (\ref{fold_Maxwell_push}).

Let $U^\Man\subset \Man$ be a generic subset and
$\phi\in\Gamma_0\Lambda^1 U^\Man$ then since $C(U^\Bman_{[i]})=\pm
U^\Man$ depending on the $\textup{sign}(\det(C_{[i]\star}))$ we have
\begin{eqnarray*}
\fl\qquad
\int_\Man \phi\wedge \hat\Jsource
&=
\DD(\hat\Jsource)[\phi]
=
\Jsource[\phi]
=
C_\varsigmaD(\Jform)
=
 \int_\Bman C^\star \phi\wedge \Jform
\\&=
\sum_{i=1}^{N(U^\Man)}
\int_{U^\Bman_{[i]}} C_{[i]}^\star \phi\wedge \Jform
=
\sum_{i=1}^{N(U^\Man)}
\int_{U^\Man} \textup{sign}(\det(C_{[i]\star}))\Fmul
C_{[i]}^{-1\star} \big(C_{[i]}^\star \phi\wedge \Jform\big)
\\&=
\sum_{i=1}^{N(U^\Man)}
\int_{U^\Man} \textup{sign}(\det(C_{[i]\star}))\Fmul
\phi\wedge C_{[i]}^{-1\star} (\Jform)
\end{eqnarray*}
since this is true for all $\phi\in\Gamma_0\Lambda^1 U^\Man$ then we have
(\ref{fold_Maxwell_pullback}).

Putting $s\colon S\hookrightarrow \Man$ and $C\colon \Bman\to\Man$ into
(\ref{AltPull_def_Q}) we see that
$\big\{(y,x)\in \Bman\times S\,\big|\ C(y)=s(x)\big\}=C^{-1}(S)$ thus
(\ref{Pull_comm_diag}) becomes
\begin{equation}
\xymatrix{
C^{-1}(S) \arhook[r]^{\hat s} \ar[d]_{\hat C}
& \Bman \ar[d]^{C}
\\
S \arhook[r]^{s} & \Man
}
\end{equation}
where for $y\in C^{-1}(S)$ we have $\hat C(y)=C(y)$ and $\hat s(y)=y$.
Hence from (\ref{fold_Maxwell_push}) and since $S$ is compact
so $1\in\Gamma_0\Lambda S$ we have 
\begin{equation*}
\fl\quad
\int_S s^\star \hat\Jsource
=
D(s^\star\hat\Jsource)[1]
=
s^\varsigma (D\hat\Jsource)[1]
=
s^\varsigma \Jsource[1]
=
s^\varsigma (C_\varsigmaD \Jform)[1]
=
\hat C_\varsigmaD (\hat s^\star \Jform)[1]
=
\int_{C^{-1}(S)} \hat s^\star \Jform
\end{equation*}
I.e. (\ref{fold_Maxwell_intg}).
\end{proof}

\begin{lemma}
Equation 
(\ref{MV_def_W}),(\ref{MV_Liou_eqn}) and (\ref{fold_rho}) 
imply
\begin{equation}
\nabla_{\Cd} \Cd = \dual{i_\Cd F}
\label{fold_Lorentz_force}
\end{equation}
\end{lemma}
\begin{proof}
From (\ref{MV_Liou_eqn}) and (\ref{fold_rho}) we have 
$i_\LiouV \Cd_\varsigmaD(\Jform)=0$. Thus from lemma
\ref{lm_intg_surf} there exists $u\in\Gamma TB$ 
such that
$\Cd_\star(u|_{(\tau,\Vy)})=\LiouV|_{\Cd(\tau,\Vy)}$. 
\begin{equation*}
\fl\qquad
C_\star(\partial_\tau|_{(\tau,\Vy)})
=
\Cd(\tau,\Vy)
=
\pi_\star\LiouV|_{\Cd(\tau,\Vy)}
=
\pi_\star\Cd_\star(u|_{(\tau,\Vy)})
=
(\pi\circ\Cd)_\star(u|_{(\tau,\Vy)})
=
C_\star(u|_{(\tau,\Vy)})
\end{equation*}
Hence $u=\partial_\tau$. Thus 
$\Cd_\star(\partial_\tau|_{(\tau,\Vy)})=\LiouV|_{\Cd(\tau,\Vy)}$.
Keeping $\Vy$ as a constant then $\Cd(\tau,\Vy)$ is an integral
curve of $\LiouV$ and hence form (\ref{Lift_force_eqn}) we have
(\ref{fold_Lorentz_force}).
\end{proof}

\subsection{The water bag model}
\label{ch_water}

Let $N\subset\Real^3$ be a bounded 3 dimensional manifold with
boundary $\iota\colon B\hookrightarrow N$.  Let $\fna\colon N\times
\Man\hookrightarrow \Ebun$ be an embedding such that $\pi\circ
\fna=\pr_2\colon N\times\Man\to\Man$ is the second projection.  This induces
the map $v=(\fna\circ\iota)\colon B\times \Man\hookrightarrow\Ebun$ so
$\pi\circ v=\pr_2$.  Let
$\alpha=\fna^\star(i_\LiouV\Omega)\in\Gamma\Lambda^6(N\times \Man)$.
We call this the \defn{water bag model} \cite{davidson}.

The source for Maxwell's equations for the water bag model is given by
\begin{equation*}
\Jsource
= \pi_\varsigma \fna_\varsigmaD \alpha
\end{equation*}
Comparing with (\ref{MV_Maxwell_Curr}), this is a regular distribution
$\Jsource=\DD(\star\tilde J)$ where $\hat\Jsource\in\Gamma \Lambda^3\Man$
\begin{equation*}
\hat\Jsource=
\Big(
\int_{\pi^{-1}(x)\inter \fna(N)}
\frac{y^a \Fmul\sqrt{|\det(\metric)|}}{y_0}\Fmul
dy^{123}\Big)\,
i_{{\partial}/{\partial x^a}} \star1
\end{equation*}

Let $v=\fna\circ\iota\colon B\times\Man\hookrightarrow\Ebun$ 
and  $v(\xi,x)=v_\xi(x)$ so for each $\xi\in B$,
$v_\xi\in\Gamma\Ebun$. There is no requirement that $v_\xi$
satisfies the cold charged fluid equations (\ref{cold_Lorentz_force}).
However we can find a reparametrisation
$w_\zeta(x)=w_{\zeta(\xi,x)}(x)=v_\xi(x)$
such that $w_\zeta$ does satisfy (\ref{cold_Lorentz_force}).

To see this, let $\Sigma\in\Man$ be a Cauchy hypersurface so that
$\pi^{-1}\Sigma\subset\Ebun$ is an initial hypersurface with respect
to $\LiouV$. Thus we have the submanifold $\sigma\colon
B\times\Sigma\hookrightarrow\Ebun$. Given any $(\zeta,y)\in
B\times\Sigma$ let $\Cd_{\sigma(\zeta,y)}\colon \Real\to\Ebun$ be the
integral curve of $\LiouV$ passing though $\sigma(\zeta,y)$, which
therefore satisfies (\ref{point_Lorentz_force}).  Given $\xi\in B$ and
$x\in\Man$, since $v_\xi(x)\in v(B\times\Man)$ then $v_\xi(x)$
lies on some integral curve $\Cd_{\sigma(\zeta,y)}$ and hence there exists
unique $\tau\in\Real$, $\zeta\in B$ and $y\in\Sigma$ such that
$v_\xi(x)=\Cd_{\sigma(\zeta,y)}(\tau)$. Let $\zeta(\xi,x)$ be the
corresponding $\zeta$ so that
$v_\xi(x)=\Cd_{\sigma(\zeta(\xi,x),y)}(\tau)$. Now set
$w_{\zeta(\xi,x)}(x)=v_\xi(x)$ so $w_\zeta$ obeys
(\ref{cold_Lorentz_force}).

\section{Conclusion and Discussion}

We have presented the distributional Maxwell-Vlasov equations and analysed
the consequences of pushforward distributional solutions. We have paid
particular attention to the question of when a pushforward
distributional solution can be the source for the Maxwell equations.

There are a number of directions that this research may continue:

It is possible to extend the idea of a submanifold distribution to
include distribution with support on submanifolds with different
dimension or intersecting components. One can ask how to extend
the notion of a pullback to these distributions.
From equation (\ref{Trans_int_soln}) we can see it is possible to take any
distribution on an initial hypersurface and create a distributional
solution to the transport equations.

Submanifold solutions with dimension of 5 or
6 may, for example, correspond to Vlasov one particle
probability functions which at every point in spacetime have a
range of possible values of one component of velocity, but fixed values
for the other components. 

As stated in the introduction, it may be possible to solve the
Maxwell-Vlasov equation numerically, for distributional solutions of
arbitrary dimension. For low dimension it will be necessary to perform
some form of regularisation of the electromagnetic field $F$, however
this should be easier for the solutions of dimension 2 or 3. These
numerical solutions can either be interpreted in their own right as,
for example, the dynamics of strings or disks, or as approximations
to regular solutions.

The stress energy tensor for a distributional source on the upper unit
hyperboloid is known. This enables one to couple the Maxwell-Vlasov
equations with Einstein's equations to give the
Einstein-Maxwell-Vlasov system \cite{Rendall}. Since the equations in
this article are already covariant no further modification would be
required.

\ack

The author would like to think all the members of the Mathematical
physics groups at Lancaster University and the Cockcroft Institute. In
particular Dr. Gabriel Bassi, Dr. David Burton, Dr. Volker Perlick,
Dr. Adam Nobel and Prof. Robin Tucker.



\end{document}